%
% Perturbative corrections to Yang-Mills partition functions
%

\input harvmac

\newcount\yearltd\yearltd=\year\advance\yearltd by 0

\noblackbox

\input epsf

\def\tilde{\widetilde}
\def\hat{\widehat}
\newcount\figno
\figno=0
\def\fig#1#2#3{
\par\begingroup\parindent=0pt\leftskip=1cm\rightskip=1cm\parindent=0pt
\baselineskip=11pt
\global\advance\figno by 1
\midinsert
\epsfxsize=#3
\centerline{\epsfbox{#2}}
\vskip 12pt
{\bf Fig.\ \the\figno: } #1\par
\endinsert\endgroup\par
}
\def\figlabel#1{\xdef#1{\the\figno}}
\def\encadremath#1{\vbox{\hrule\hbox{\vrule\kern8pt\vbox{\kern8pt
\hbox{$\displaystyle #1$}\kern8pt}
\kern8pt\vrule}\hrule}}

\def\half{{\textstyle{1\over2}}}

\def\half{{1\over 2}}

 \def\d{{\delta}}
 
 \def\t{{\theta}}
 \def\a{{\alpha}}
 
 \def\frac#1#2{{#1\over #2}}
 \def\l{{\lambda}}
 
 \def\D{{\Delta}}
 \def\g{{\gamma}}
 \def\s{{\sigma}}
 
 \def\b{{\beta}}
 \def\t{{\tau}}
 
 \def\CC{{\cal C}}

 \def\CD{{\cal D}}
 \def\p{\partial}

 \def\r{\rightarrow}
\def\bal{{\bar{\alpha}}}
\def\bb{{\bar{\beta}}}
\def\bg{{\bar{\gamma}}}
\def\bd{{\bar{\delta}}}
\def\bl{{\bar{\lambda}}}
\def\br{{\bar{\rho}}}

\def\r{{\rho}}
\def\bs{{\bar{\sigma}}}
\def\bt{{\bar{\tau}}}
\def\D{\nabla}
\def\d{\partial}
%%%%%%%%%%% References %%%%%%%%%%%%%%%%%%%%%%%%%%%%%%%%%%%%%%%

\lref\grosswit{
D.~J.~Gross and E.~Witten,
``Possible Third Order Phase Transition In The Large N Lattice Gauge Theory,''
Phys.\ Rev.\ D {\bf 21}, 446 (1980).
%%CITATION = PHRVA,D21,446;%%
}

\lref\creutz{
M.~Creutz,
``On Invariant Integration Over SU(N),''
J.\ Math.\ Phys.\  {\bf 19}, 2043 (1978).
%%CITATION = JMAPA,19,2043;%%
}

%\AharonySX
\lref\first{
O.~Aharony, J.~Marsano, S.~Minwalla, K.~Papadodimas and M.~Van Raamsdonk,
``The Hagedorn / deconfinement phase transition in weakly coupled large N gauge
theories,''
arXiv:hep-th/0310285.
%%CITATION = HEP-TH 0310285;%%
}

%\AharonyBQ
\lref\second{
  O.~Aharony, J.~Marsano, S.~Minwalla, K.~Papadodimas and M.~Van Raamsdonk,
  %``A first order deconfinement transition in large N Yang-Mills theory on  a
  %small S**3,''
  Phys.\ Rev.\ D {\bf 71}, 125018 (2005)
  [arXiv:hep-th/0502149].
  %%CITATION = HEP-TH 0502149;%%
}

\lref\sundborg{
  B.~Sundborg,
  %``The Hagedorn transition, deconfinement and N = 4 SYM theory,''
  Nucl.\ Phys.\ B {\bf 573}, 349 (2000)
  [arXiv:hep-th/9908001].
  %%CITATION = HEP-TH 9908001;%%
}

\lref\lattice{
M.~Teper,
  %``The Finite temperature phase transition of SU(2) gauge fields in
  %(2+1)-dimensions,''
  Phys.\ Lett.\ B {\bf 313}, 417 (1993).
  %%CITATION = PHLTA,B313,417;%%

J.~Christensen, G.~Thorleifsson, P.~H.~Damgaard and J.~F.~Wheater,
  %``Thermodynamics of SU(3) lattice gauge theory in (2+1)-dimensions,''
  Nucl.\ Phys.\ B {\bf 374}, 225 (1992).
  %%CITATION = NUPHA,B374,225;%%

Philippe~ de~ Forcrand,~ Oliver Jahn,
`` Deconfinement transition in 2+1-dimensional SU(4) lattice gauge theory,''
hep-lat/0309153

K.~Holland,
``Another weak first order deconfinement transition: three-dimensional SU(5) gauge theory,''
JHEP\ 0601 (2006) 023,

J.~ Liddle, M.~ Teper
 ``The deconfining phase transition for SU(N) theories in 2+1 dimensions''
arXiv:hep-th/0509082.
}

\lref\integrals{
This file is available at http://www.fas.harvard.edu/~papadod/3loop/3loop.html.}

\lref\sy{
B.~Svetitsky, L. ~G.~ Yaffe ,
``Critical Behavior At Finite Temperature Confinement Transitions,''
Nucl.\ Phys.\ B210:423,1982 .
}

\lref\dk{
  M.~R.~Douglas and V.~A.~Kazakov,
  %``Large N phase transition in continuum QCD in two-dimensions,''
  Phys.\ Lett.\ B {\bf 319}, 219 (1993)
  [arXiv:hep-th/9305047].
  %%CITATION = HEP-TH 9305047;%%
}

\lref\dhoker{
  E.~D'Hoker,
  %``Perturbative Results On QCD In Three-Dimensions At Finite Temperature,''
  Nucl.\ Phys.\ B {\bf 201}, 401 (1982).
  %%CITATION = NUPHA,B201,401;%%
}

\lref\order{
  D.~V.~Boulatov,
  %``Wilson loop on a sphere,''
  Mod.\ Phys.\ Lett.\ A {\bf 9}, 365 (1994)
  [arXiv:hep-th/9310041].
  %%CITATION = HEP-TH 9310041;%%
}

\lref\largenlattice{
R.~Narayanan and H.~Neuberger,
  %``Infinite N phase transitions in continuum Wilson loop operators,''
  JHEP {\bf 0603}, 064 (2006)
  [arXiv:hep-th/0601210].
  %%CITATION = HEP-TH 0601210;%%

  F.~Bursa and M.~Teper,
  %``Strong to weak coupling transitions of SU(N) gauge theories in 2+1
  %dimensions,''
  arXiv:hep-th/0511081.
  %%CITATION = HEP-TH 0511081;%%
}

\lref\wittenads{
  E.~Witten,
  ``Anti-de Sitter space, thermal phase transition, and confinement in  gauge
  theories,''
  Adv.\ Theor.\ Math.\ Phys.\  {\bf 2}, 505 (1998)
  [arXiv:hep-th/9803131].
  %%CITATION = HEP-TH 9803131;%%
}

\lref\sva{
Edmonds, A. R.  Angular momentum in quantum mechanics  Princeton University Press, 1974.

}
\lref\svb{
Jones, M. N. (Michael Norman), Spherical harmonics and tensors for classical field theory 
Published: Letchworth, Hertfordshire, England : Research Studies Press ; New York : Wiley, c1985.

}

%%%%%%%%%%%%%%% Title Page %%%%%%%%%%%%%%%%%%%%%%%%%%%%%%%%%%%%

\Title
{\vbox{\baselineskip12pt
\hbox{hep-th/0612066}
\hbox{ITFA-2006-50}}}
{\vbox{\centerline{A second order deconfinement transition for large $N$} \vskip 10pt \centerline{2+1 dimensional Yang-Mills theory on a small $S^2$}}}

\centerline{ Kyriakos Papadodimas$^{a,b}$, Hsien-Hang Shieh$^{c}$ and Mark Van Raamsdonk$^{c}$ }

\medskip

\centerline{\sl $^{a}$Jefferson Physical Laboratory, Harvard University,Cambridge, MA 02138, USA}
\centerline{\sl $^{b}$Institute for Theoretical Physics,
University of Amsterdam}
\centerline{\sl
Valckenierstraat 65,
1018 XE Amsterdam,
The Netherlands}

\centerline{\sl $^{c}$Department of Physics and Astronomy, University of British Columbia}
\centerline{\sl Vancouver, BC, Canada, V61 1Z1}

\medskip

\vskip 0.5cm

\centerline{\bf Abstract}
\medskip
\noindent
We study the thermodynamics of large $N$ pure 2+1 dimensional Yang-Mills theory on a small spatial $S^2$. By studying the effective action for the Polyakov loop order parameter, we show analytically that the theory has a second order deconfinement transition to a phase where the eigenvalue distribution of the Polyakov loop is non-uniform but still spread over the whole unit circle. At a higher temperature, the eigenvalue distribution develops a gap, via an additional third-order phase transition. We discuss possible forms of the full phase diagram as a function of temperature and sphere radius. Our results together with extrapolation of lattice results relevant to the large volume limit imply the existence of a critical radius in the phase diagram at which the deconfinement transition switches from second order to first order. We show that the point at the critical radius and temperature can be either a tricritical point with universal behavior or a triple point separating three distinct phases.

\vskip 0.5cm
\Date{December 2006}
%\listtoc
%\writetoc

%%%%%%%%%%%%%%%%%% Body of paper %%%%%%%%%%%%%%%%%%%%%%%%%%%%%%%%%%%%

%\draftmode

\newsec{Introduction}

In this note, we follow \second\ to study the thermodynamics of large $N$ pure 2+1 dimensional Yang-Mills theory on a spatial $S^2$ with radius much smaller than the scale set by the dimensionful coupling of the gauge theory. In this limit, the dimensionless coupling $\lambda = g^2 N R$ is small, so the thermodynamics can be studied in perturbation theory. The thermal partition function is given by the path integral for the Euclidean theory on $S^2 \times S^1$, where the $S^1$ has radius given by the inverse temperature $\beta = T^{-1}$. From this path integral expression, we integrate out all modes apart from the trace of the Polyakov loop,\foot{Here $A$ is averaged over the $S^2$.}
$$
u = {1 \over N} \tr(U) = {1 \over N} \tr P e^{i \int_{S^1} A} \; ,
$$
giving us an effective action for $u$, the standard order parameter for confinement. This effective action takes the form (valid for $|u| \le {1/2}$)
\eqn\seff{
S_{eff}(u) = f(T, \lambda) |u|^2 + \lambda^2 b(T) |u|^4 + {\cal O}(\lambda^4) \; .
}
For low temperatures, $f$ is positive, and so the saddle-point configuration has $u=0$. At some temperature $T_H = T_H^0 + {\cal O}(\lambda)$
, $f$ becomes negative, so $u=0$ is no longer the minimum action configuration. The three-loop calculation in this paper shows that the coefficient $b$ is positive at the critical temperature, so as $f$ becomes negative, $|u|$ develops an expectation value gradually, resulting in a second-order phase transition.

The situation here is in contrast to the 3+1 dimensional case, where the analogous calculation \second\ revealed a negative value for $b(T_H)$. In that case, the deconfinement transition is first order, characterized by a discontinuous change in the eigenvalue distribution for the unitary matrix $U$ from the uniform distribution (eigenvalues equally distributed around the unit circle) to a non-uniform distribution with a gap (i.e. a region of the circle where no eigenvalues are present). In terms of the eigenvalue distribution, the continuous transition we find here corresponds to a continuous change from the uniform distribution to an increasingly non-uniform distribution. As argued in \first, at a temperature
$$
T_2 = T_H + {1 \over 2} \lambda^2 b(T_H)/f'(T_H)
$$
the minimum value of the eigenvalue density will reach zero, as depicted in figure 1, and the eigenvalue distribution will develop a gap for higher temperatures. This results in an additional third-order phase transition of Gross-Witten type, so the phase diagram for pure Yang-Mills theory on $S^2 \times S^1$ contains at least three distinct phases (uniform, non-uniform, gapped).

\fig{Distribution of Polyakov loop eigenvalues on unit circle (horizontal axis) in confined phase (a), in gapless phase above second order deconfinement transition (b), at third-order gapping transition (c), and in high-temperature gapped phase (d).}{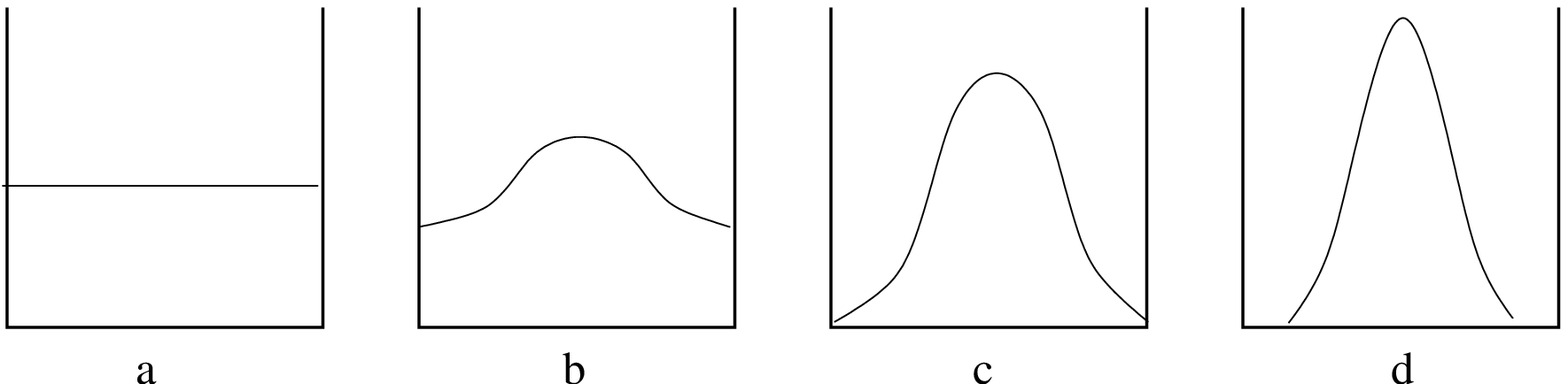}{6.0truein}

The basic setup for our calculation and the calculation itself are presented in sections 2 and 3 of this paper. We then consider the implications of our results for the full phase diagram as a function of temperature and spatial radius. In section 4, we argue that in the high-temperature limit for fixed radius the theory effectively reduces to pure two-dimensional Yang-Mills theory on a spatial $S^2$. This theory has a third order phase transition as the radius of the sphere is varied $\dk$, suggesting that our phase diagram has an additional third order transition line coming from large temperature along the curve
$$
TR \sim 1 / (\lambda R)  \; .
$$
It is tempting to conjecture that this third-order transition line connects up with the one emerging from the critical temperature at small radius (as in figure 6b). Indeed, these are qualitatively very similar: while the latter transition is associated with the development of a gap in the eigenvalue distribution for the Polyakov loop, the two-dimensional Yang-Mills theory transition is associated with the development of a gap in the eigenvalue distribution for the Wilson loop around the equator of the sphere (or any other maximal-area non-intersecting loop)$\order$. On the other hand, we have not been able to show any direct relation between the two order parameters (which we can choose to be the minimum value of the eigenvalue density for these two Wilson loops). Further, it is possible that the high temperature transition ceases to be sharp as we decrease the temperature.

In section 5, we consider the fate of the actual deconfinement transition line as the radius is increased and the possible forms for the full phase diagram. At large volume, we can appeal to lattice results, which for finite $N$ suggest that there is a second order deconfinement transition for $N=2,3$ (and possibly $N=4$), a weak first order transition for $N=5$ and a stronger first-order transition for $N=6$ \lattice. It is believed that the transition should continue to strengthen as $N$ is increased. If this is correct, the large $N$ theory would be expected to exhibit a first order deconfinement transition, and there must be a point along the deconfinement transition line at some critical radius, where the deconfinement transition switches from second order to first order. We argue that there are two qualitatively different behaviors of the Polyakov loop effective potential that can lead to such a point. These correspond either to a tricritical point at which both quadratic and quartic terms in an effective action vanish, or to a triple point, separating three distinct phases. The simplest possible phase diagrams consistent with the available information are presented in figure 6.

We conclude in section 6 with a few comments on the implications for a possible gravitational dual theory of the existence of three distinct phases.

\newsec{The set up}
In this section we will briefly review the results and techniques
developed in \second, \sundborg, \first\ and apply them to analyze the
thermodynamic properties of $2+1$ dimensional,
large $N$ pure Yang-Mills theory living on
a two-sphere of radius $R$. As argued in \first, asymptotically free gauge theories become weakly coupled at small volume, in this case when $\lambda = g^2 N R \ll 1$. This is
because all modes with the exception of the constant mode of $A_0 $ on $S^2 \times S^1$ (which
we denote by $\alpha$), are massive, with masses $\geq 1/R$. We thus have an effective IR
cutoff on the renormalization group flow of the coupling $g_{YM}^2 N$. The massive modes may be integrated out directly in perturbation theory, yielding an effective action for the zero-mode $\alpha$, or more precisely, for the constant $SU(N)$ matrix $U=e^{i\beta\alpha}$. The result of \first\ shows that in the large $N$ limit, the matrix model undergoes a Hagedorn-like transition at some critical temperature, which corresponds to a deconfinement transition in the gauge theory. We note that the large $N$
limit is essential for the theory on a compact manifold to exhibit
a sharp phase transition.

\subsec{Basic setup}

The thermal partition function of pure $SU(N)$ Yang-Mills theory on $S^2$
at temperature $T$ can be evaluated by the Euclidean path integral
of the theory on the manifold $S^2 \times S^1$, with the radius of
the circle $S^1$ equal to $\beta={1\over T}$. The action of the
theory is: \eqn\pureymlag{{\cal L} = {1\over 4} \int_0^\beta dt \int
d^2x \tr(F_{\mu \nu} F^{\mu \nu}).} To perform this computation we
will work in the gauge: \eqn\gf{\del_i A^i=0} where $i=1, 2$ runs
over the sphere coordinates, and $\del_i $ are (space-time)
covariant derivatives.

This does not completely fix the gauge, as we can still make
spatially independent gauge transformations. To completely fix the
gauge we also impose \eqn\gff{\p_t \int_{S^2} A_0=0.} so we choose
the constant (on the sphere) mode of $A_0$ to be independent of
time. We define: \eqn\defalpha{\alpha = {g_{YM}\over \omega_2}
\int_{S^2} A_0,} where $\omega_2$ is the volume of the 2-sphere.

The mode $\alpha$ is a zero mode of the theory which is always
strongly coupled and cannot be integrated out perturbatively.
Because of this, we will compute the path integral of the theory in
two steps. First we will integrate out all other modes which are
massive to generate an effective action for the zero mode
$\alpha$. Then we will analyze the effective action for $\alpha$.
In other words, we will do the path integral in the following order:

\eqn\pathintegral{Z = \int DA\, d\alpha \,e^{-S[A,\alpha]} = \int
d\alpha \,\int dA\, e^{-S[A,\alpha]} = \int d\alpha\,
e^{-S_{eff}[\alpha]}}

As explained in $\first$ the effective action for $\alpha$ can be
written completely in terms of the unitary matrix $U\equiv e^{i
\beta \alpha}$ in the form:

\eqn\wexplia{ \eqalign{S_{eff}=&\sum_{m} C_{m,-m} Tr (U^m ) Tr
(U^{-m})+ \lambda \beta \sum_{m,n}C_{m,n,-m-n}Tr (U^m ) Tr (U^n )
Tr(U^{-m-n})/N\cr
 +&\lambda^2 \beta \sum_{m,n,p} C_{m,n,p,-m-n-p} Tr (U^m ) Tr (U^n ) Tr(U^{p}) Tr(U^{-m-n-p})/ N^2 + \dots }
} Then the free energy of the theory is given by the matrix
integral:

\eqn\matrixint{Z(\beta)=e^{-\beta F} = \int [dU] e^{-S_{eff}(U)}}
where, as discussed in $\first$, the Fadeev-Popov determinant of the
gauge fixing transforms the integral over $\alpha$ to an integral
over the gauge group with Haar measure $[dU]$. Note $U$ is the holonomy of the
gauge field around the temporal circle. We thus have an effective
action for the temporal Wilson loop, whose expectation value is the standard order parameter for the deconfinement transition.

In the large $N$ limit, we can evaluate the integral using saddle
point techniques. Introducing the eigenvalue distribution
$\rho(\theta )=\sum_i \delta (\theta -\theta _{i})/N$ where $\theta_
i $ are the eigenvalues of $\alpha$, $i=1..N$ and defining $u_n
=\int d\theta \rho (\theta)e^{in\theta} =Tr(U^n )/N $ the effective
action takes the form: \eqn\wmomen{ Z[\beta]=\int [du_n ] [d\bar{u}_n
]e^{-N^2 S'_{eff}[u_n, \bar{u_n }; \beta, \lambda]} } with
\eqn\zz{\eqalign{ S'_{eff}=&\sum_{m} (1/m-C_{m,-m}) u_m \bar{u_m }+
\lambda \beta \sum_{m,n} C_{m,n,-m-n} u_m u_n \bar{u_{m+n}}\cr
+&\lambda^2 \beta \sum_{m,n,p} C_{m,n,p,-m-n-p} u_m u_n u_p
\bar{u_{m+n+p}} + ... }} where the extra $1/m$ comes from
the Vandermonde determinant obtained in going to the variables $u_n$.

Note that $u_{n}=0$ is a stationary point at all temperatures. It
corresponds to the uniform distribution of the eigenvalues of $U$.
The stability of this saddle point depends on the values of the
coefficients $C_{n,m,...}$. As we will show in the next subsection,
the coefficient of $|u_1|^2$ is positive at small $T$ but becomes
negative as we increase the temperature. This signals that the
$u_{n}=0$ phase becomes unstable and the system undergoes a phase
transition. As explained in $\first$, near the transition
temperature the $u_{1}$ mode becomes massless, while the $u_{n>1}$ modes
remain massive. We will further integrate out these higher moments
to obtain an effective action for $u_{1}$ near $T_H $ to analyze the
order of the phase transition. To achieve this we will need the
coefficient of the quartic term in $u_{1}$. The relevant terms in
$\zz $ are: \eqn\zza{\eqalign{ S'_{eff}=& (1-C_{1,-1})  |u_1 |^2+
(1/2-C_{2,-2}) |u_2 |^2 + ...\cr
         +&\lambda \beta [ C_{1,1,-2} (u_1 ^2 \bar{u_{2}}+ u_2 \bar{u_{1}}^2  )+ ... ]\cr
         +&\lambda^2 \beta  C_{1,1,-1,-1} |u_{1}| ^4 + ...
}}
For the saddle point configuration, the higher modes $u_n$ are determined by minimizing the effective action over $u_n$ for fixed $u_1$, so we have
\eqn\zca{u_{2}= - \lambda
\beta \frac{C_{1,1,-2}}{(1/2-C_{2,-2}) }u_1
^2 + O(\lambda^2 ) }
This gives the effective action:
\eqn\zqa{
S'_{eff}(u_1)= (1-C_{1,-1})  |u_1 |^2 +\lambda^2 \beta  b_{c} |u_{1}|^4 +
... }
where
\eqn\bc{ b_c = - \beta ^2 C_{1,1,-2}(\beta )^2/
(1/2-C_{2,-2}(\beta ))+ \beta C_{1,1,-1,-1}+ O(\lambda  ) }
As discussed in $\first$, the sign of $b$ evaluated at the critical temperature
determines the order of the phase transition.  If $b_c <0$ the transition is first order and
occurs at the Hagedorn temperature $T_H$ at zero coupling, but
slightly below $T_H$ at small but finite coupling (they are the
same up to order $\lambda ^2 $). If $b_c >0$ there are two phase
transitions. The first one occurs at exactly $T_H $ and is second
order, while the second one is a Gross-Witten type third order phase
transition happening above $T_H $. The third order phase transition
occurs at the point where the eigenvalue distribution of $U $
develops a gap. Our goal will be to compute the coefficients
$C_{n,m,...}$ to the appropriate order in perturbation theory to
obtain $b_c $. The $C_{2,-2}$ requires a one loop calculation, while
$C_{1,1,-2}$ and  $C_{1,1,-1,-1}$ require two and three loop
calculations respectively.

\subsec{One loop free energy}

It is shown in $\first$ that the coefficient $C_{m,-m}$ can be
extracted by computing the one loop partition function of the theory
. It turns out that: \eqn\ccm{
 C_{m,-m} =-\frac{ z_V (x^m ) }{m}
} where $x=e^{-\beta/R}$ and $z_V (x )$ is the single particle
partition function \eqn\slp{ z_{V}(x) = \sum_\triangle  n(\triangle
)x^{-\beta \triangle } } In our case  $\triangle ^2 $ are the
eigenvalues of the Laplacian on the unit two sphere acting on the
vector spherical harmonics surviving the gauge fixing, while
$n(\triangle )$ is the multiplicity of each mode. More specifically
$\triangle_h ^2 =h( h+1)\,$, $n(\triangle_h )=2h+1$, $h=1,2,...$ The
Hagedorn temperature is determined by the point where the $u_{1}$
mode become unstable: \eqn\hag{ z_{V}(x_c) =1 } We are not able to get
a closed form for $z_V (x )$, but numerically we find:

\eqn\hagtemp{x_c\simeq1.195097} so the Hagedorn temperature of the
free theory on a two-sphere of radius $R=1$ is: \eqn\hagetempb{T_H\simeq 0.302675}

Similarly,

\eqn\ccc{
 C_{2,-2} =-\frac{ z_V (x_c ^2 ) }{2}= 0.38155
}

\subsec{Gauge fixed action}

Formally, the thermal partition function is of the form \eqn\parti{ Z[\beta
]=\int [d\alpha ] [d A'_0 ] [d A_i ] \triangle _1 \triangle _2
e^{-S[\alpha , A'_0 , A_i ; \beta ]} } where $\triangle_1 $,
$\triangle_2 $ are the Fadeev-Popov determinants associated with
$\gf , \gff$  respectively and $\beta = 1/T$ is the size of the time
circle. The prime on $ A'_0 $ signals that the constant mode has
been removed. As shown in $\first$, $\triangle_2 $ can be combined
with $\left[ d \alpha \right]$ to give the integration measure of a
unitary matrix $[d U ]$ with $U=e^{i\beta\alpha }$. On the other hand,
$\triangle_1 $ is \eqn\fpd{ \det{\del_i D^i } = \int \CD c \CD {\bar
c} e^{- \tr({\bar c} \del_i D^i c)}} where $D^i$ denotes a gauge
covariant derivative \eqn\gcovar{D_i c=\del_i-i g_{YM} [A_i,c]} and
$c$ and ${\bar c}$ are complex ghosts in the adjoint representation
of the gauge group. With our gauge choice, the quadratic terms in
the Yang-Mills action \pureymlag\ take the form \eqn\fymact{ -\int
d^3x \tr \left( \half A_i (\tilde{D}_0^2 +\del^2) A^i + \half A'_0
\del^2 A'_0 + {\bar c} \del^2 c\right)} where \eqn\deft{\tilde{D}_0
X \equiv \p_0 X -i[\alpha, X].} The interaction terms in \pureymlag\
are given by \eqn\ymint{\eqalign{ \int d^3x \tr (  & ig_{YM}
\tilde{D}^0 A^i [A_i,A'_0] -ig_{YM}[A^i,A'^0] \del_i A'_0 -i
g_{YM}\del_i A_j [A^i, A^j] + \cr & {g_{YM}^2\over 4} [A_i,
A_j][A^j, A^i] -{g_{YM}^2\over 2} [A'_0,A_i][A'^0,A^i] -i
g_{YM}\del_i {\bar c} [A_i, c]). \cr}} We wish to obtain the
effective action by \eqn\wact{ e^{-S_{eff}[U; \beta,
\lambda]}=\int [d c ] [d A'_0 ] [d A_i ] e^{-S_{1loop}}<
e^{-S_{int}} >. } The result is an effective theory for a constant
$SU(N)$ matrix $U=e^{i\beta\alpha}$ \eqn\w{ Z[\beta]=\int [dU]
e^{-S_{eff}[U; \beta, \lambda]}. }

\newsec{The Perturbative Calculation}
In this section, we summarize the computation of the two and three loop diagrams. As we are on $S^2$, it will be convenient to expand
the fields in terms of the vector and scalar spherical harmonics on $S^2 $.  As a result, the spatial momentum
integrals in the Feynman diagrams can be replaced by sums over the quantum numbers of the generators of $SU(2 )$.
We will first set up the conventions that facilitate the computation.
\subsec{Spherical Harmonics Expansion on $S^2 $}
The basic set-up for the computation was described in section 2 of \first. As in the $3+1$ case, it will be useful to write the action explicitly in terms of a set of spherical harmonic integrals. We will denote the scalar and
vector spherical harmonics on $S^2$ by $S^\alpha(\theta)$ and
$V_i^\beta(\theta)$, where
$\alpha = (j_\alpha ,m_\alpha)$ and $\beta = (j_\beta,
m_\beta)$ are the $SU(2 )$ quantum numbers for the various modes. Our conventions for the vector spherical harmonics can be found in appendix A. Note for $S^{j,m}$ and
$V_i^{j,m}$ the $j$ quantum number starts at $j=0$, $j=1$, respectively.
In terms of these, we have:
\eqn\expand{\eqalign{
A'_0(t,\theta) & = \sum_{\alpha} a^{\alpha}(t) S^{\alpha}(\theta); \cr
A_i(t,\theta) &= \sum_{\beta} A^{\beta}(t) V_i^{\beta}(\theta); \cr
c(t,\theta) &= \sum_{\alpha} c^{\alpha}(t) S^{\alpha}(\theta). \cr}}
We will denote the complex conjugate of $S^{\alpha}$ by $S^{\bar \alpha}$.
On $S^2$, the complex conjugation for scalar spherical harmonics
is the same as inverting the sign of $m_{\alpha}$
and multiplying by $(-1)^{m_{\alpha}}$.
We also denote the complex conjugate of $V_i^{\beta }$ by $V_i^{\bar \beta }$.
The complex conjugation for vector spherical harmonics
has the effect of changing the sign of $m_{\beta}$
and multiplying by $(-1)^{m_{\beta}+1}$. The scalar spherical
harmonics are an orthonormal basis of functions on $S^2$,
\eqn\ortho{\int_{S^2} S^{\alpha} S^{\bar \beta} = \delta^{\alpha \beta}.}
In terms of these spherical harmonics, we can define
\eqn\sphconv{\eqalign{
C^{\alpha \beta \gamma} &= \int_{S^2} S^\alpha \vec{V}^\beta \cdot \vec{\nabla} S^\gamma, \cr
D^{\alpha \beta \gamma} &= \int_{S^2} \vec{V}^\alpha \cdot \vec{V}^\beta S^\gamma, \cr
E^{\alpha \beta \gamma} &= \int_{S^2} (\vec{\nabla} \times \vec{V}^\alpha )\cdot (\vec{V}^\beta \times \vec{V}^\gamma )
\cr}}
They appear as effective couplings of various interaction terms in the pure Yang Mills Lagrangian. These integrals can be computed explicitly using properties of $SU(2 )$. We will list the expressions for $C$, $D$ and $E$ in the appendix. Note that $C$ is antisymmetric in $\alpha$ and $\gamma$, $D$
is symmetric in $\alpha$ and $\beta$, and $E$ is antisymmetric in $\beta$ and $\gamma$.
Using these expressions, the quadratic part of the action for pure
Yang Mills theory on the two sphere becomes
\eqn\qua{
S_2 =
\int dt \tr \left( {1 \over 2} A^{\bar \alpha} (-D_\tau ^2 + j_\alpha (j_\alpha +1) ) A^\alpha + {1 \over 2}a^{\bar{\alpha}}j_\alpha (j_\alpha +1)a^\alpha + \bar{c}^{\bar{\alpha}}j_\alpha (j_\alpha +1)c^\alpha \right).
}
The cubic interactions are
\eqn\cubic{\eqalign{
S_3 = g_{YM}
\int dt \tr( & i \bar{c}^{\bar \alpha} [A^\gamma, c^\beta] C^{\bar{\alpha}
\gamma \beta} +2i a^\alpha A^\gamma a^\beta
C^{\alpha \gamma \beta} \cr
&-i[A^\alpha, D_\tau A^\beta]a^\gamma D^{\alpha \beta \gamma} - i
 A^\alpha A^\beta A^\gamma E^{\alpha \beta \gamma}),
}} 
and the quartic interactions are given by \eqn\quartic{\eqalign{ S_4
= g_{YM}^2 \int dt \tr( & -{1 \over 2} [a^\alpha, A^\beta][a^\gamma,
A^\delta]\left( D^{\beta \bar{\lambda} \alpha} D^{\delta \lambda
\gamma} + {1 \over j_\lambda (j_\lambda+1)} C^{\alpha \beta
\bar{\lambda}} C^{\gamma \delta \lambda} \right) \cr & - {1 \over 2}
A^\alpha A^\beta A^\gamma A^\delta \left( D^{\alpha \gamma
\bar{\lambda}} D^{\beta \delta \lambda} - D^{\alpha \delta
\bar{\lambda}} D^{\beta \gamma \lambda} \right)).}}

\subsec{Effective Vertices} 
Since the action is quadratic in $a$
and $c$, these may be integrated out explicitly to give additional
effective vertices. As discussed in detail in $\second$, we have two types
of effective vertices. The A type vertices arise from loops of $a$
and $c$. The B type vertices are from open strings of $a$'s
containing two vertices linear in $a$ and some number of quadratic $a$
vertices. Both involve divergences proportional to $\delta(0)$. As
for the 3+1 dimensional theory on $S^3$,  all divergent
contributions proportional to $\delta(0)$ cancel. In particular the
contributions from the $a$ and $c$ loops completely cancel out with
the divergent parts in the B type effective vertices. We thus have
only the non-divergent contributions from the latter.
Computationally, we will keep only the vertices appearing in the
Lagrangian above which contain no temporal component $a $ and
ghosts $c$. In addition, we will also have the second type of
effective vertices, but can ignore any contributions proportional to $\delta(0)$  arising from contractions of $\D_\tau A$s in these. The B type vertices are:
\eqn\Bverts{\eqalign{ B_4 &= {g_{YM}^2 \over 2} {D^{\alpha_1 \beta_1
\gamma} D^{\alpha_2 \beta_2 \bar{\gamma}} \over j_\gamma (j_\gamma +
1)} \tr([A^{\alpha_1}, D_\tau A^{\beta_1}] [A^{\alpha_2}, D_\tau
A^{\beta_2}]), \cr B_5 &= -i g_{YM}^3 {D^{\alpha_1 \beta_1 \lambda}
C^{\bar{\lambda} \gamma \bar{\sigma}} D^{\alpha_2 \beta_2 \sigma}
\over j_\lambda (j_\lambda + 1) j_\sigma (j_\sigma + 1)}
\tr([A^{\alpha_1}, D_\tau A^{\beta_1}][A^\gamma, [A^{\alpha_2},
D_\tau A^{\beta_2}]]), \cr B_6 &= {g_{YM}^4 \over 2} \left( 3
{D^{\alpha_1 \beta_1 \sigma} C^{\bar{\sigma} \gamma_1 \tau}
C^{\bar{\tau} \gamma_2 \lambda}  D^{\alpha_2 \beta_2 \bar{\lambda}}
\over j_\lambda (j_\lambda + 1) j_\sigma (j_\sigma + 1) j_\tau
(j_\tau + 1) } + {D^{\alpha_1 \beta_1 \sigma} D^{\gamma_1 \lambda
\bar{\sigma}} D^{\gamma_2 \bar{\lambda} \bar{\tau}} D^{\alpha_2
\beta_2 \tau} \over j_\sigma (j_\sigma + 1) j_\tau (j_\tau + 1) }
\right) \cr & \qquad \tr([[A^{\alpha_1}, D_\tau
A^{\beta_1}],A^{\gamma_1}] [[A^{\alpha_2}, D_\tau
A^{\beta_2}],A^{\gamma_2}]). \cr }}

\fig{The diagrams contributing to the free energy up to 3-loop
order. In this figure we present a particular planar form for each
diagram, but in some cases the same diagram may also be drawn in
the plane in different ways.
}{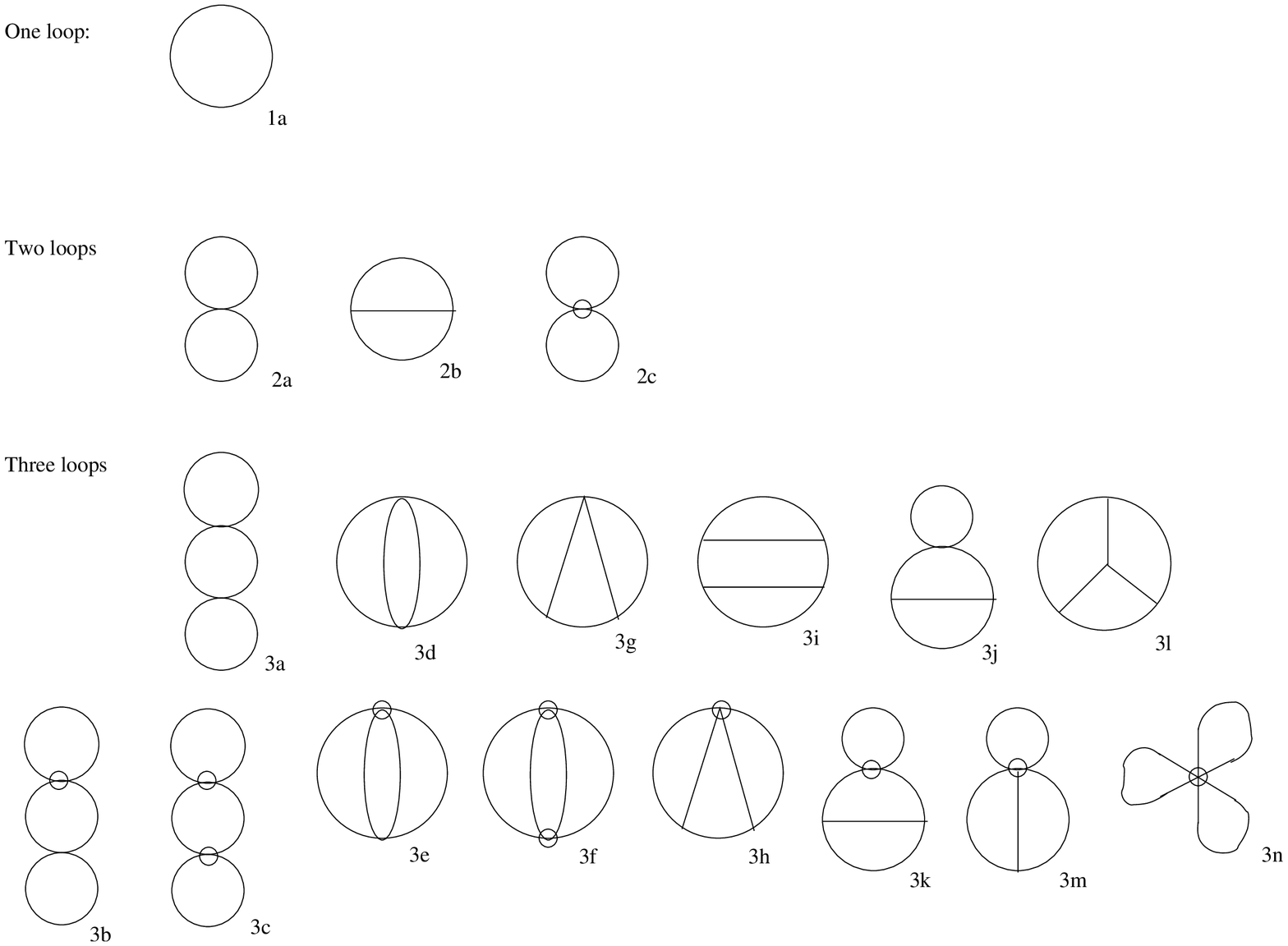}{7truein} \figlabel{\diagrams}

The relevant diagrams
contributing to the vacuum energy
at one, two and three loops are shown in figure \diagrams.
They are the same as those in 3+1 Yang-Mills theory on $S^3$. The B
type vertices are denoted by circles.

\subsec{Propagators} The propagators from the quadratic part of the
action are \eqn\propc{ \langle \bar{c}^{\bar \alpha}_{ab}(t)
c_{cd}^\beta(t') \rangle =  {1 \over L_\alpha } \delta^{\alpha
\beta} \delta(t - t') \delta_{ad} \delta_{cb}, } \eqn\propa{ \langle
a_{ab}^\alpha(t) a_{cd}^\beta(t') \rangle =  {1 \over L_\alpha }
\delta^{\alpha \bar{\beta}} \delta(t - t') \delta_{ad} \delta_{cb},
} \eqn\propbiga{ \langle A^\alpha_{ab}(t) A^\beta_{cd}(t') \rangle =
\delta^{\alpha \bar{\beta}} \Delta^{ad,cb}_{j_\alpha}(t-t'), }
\eqn\propdta{ \langle D_\tau A^\alpha_{ab}(t) A^\beta_{cd}(t')
\rangle = - \langle A^\alpha_{ab}(t) D_\tau A^\beta_{cd} (t')
\rangle = \delta^{\alpha \bar{\beta}} D_\tau
\Delta^{ad,cb}_{j_\alpha}(t-t'), } \eqn\dtdt{ \langle D_\tau
A^\alpha_{ab}(t) D_\tau A^\beta_{cd}(t') \rangle = \delta^{\alpha
\bar{\beta}} \delta(t - t') \delta_{ad} \delta_{cb} - \delta^{\alpha
\bar{\beta}} L_\alpha ^2 \Delta^{ad,cb}_{j_\alpha}(t-t'), } where
the (adjoint $\otimes$ adjoint)-valued function $\Delta_{j_a }(t)$
solves \eqn\propoeq{ (-D_\tau ^2 + j_\alpha (j_\alpha +1) )
\Delta_{j_\alpha }(t)= \delta_{j_\alpha }(t) } explicitly on
$[0,\beta)$, \eqn\defdelta{ \Delta_{j_a }^{ad, cb }(t) = {e^{i
\alpha t} \over 2L_a } \left( {e^{-L_a t} \over 1 - e^{i \alpha
\beta} e^{-L_a \beta}} -  {e^{L_a t} \over 1 - e^{i \alpha \beta}
e^{L_a \beta}} \right) } where $L_a =(j_a (j_a +1))^{1/2} $ and
$\alpha$ is short for $\alpha^{ad } \otimes 1^{ cb } - 1^{ad }
\otimes \alpha^{cb }$. The quantity that will appear in computation
is $\Delta_{j_a }(t_1-t_2)$ where $t_1$ and $t_2$ range between $0$
and $\beta$ so that the argument of $\Delta_{j_a }(t)$ takes value
between $-\beta $ and $\beta$. What we will use in computation are
the full propagators: \eqn\fulproa{ \Delta_{j_a
}(t)=\Theta(t)\frac{1}{2L_a }(\frac{e^{(i\alpha-L_a
)t}}{1-e^{i\alpha\beta }e^{-L_a\beta }}+\frac{e^{(i\alpha+L_a
)(t-\beta )}}{1-e^{-i\alpha\beta }e^{-L_a \beta
}})+\Theta(-t)\frac{1}{2L_a }(\frac{e^{(i\alpha -L_a )(t+\beta
)}}{1-e^{i\alpha\beta }e^{-L_a \beta }}+\frac{e^{(i\alpha +L_a
)t}}{1-e^{-i\alpha\beta}e^{-L_a \beta}}) }
\eqn\fulprob{D_{t}\Delta_{j_a
}(t)=\Theta(t)\frac{1}{2}(-\frac{e^{(i\alpha -L_a
)t}}{1-e^{i\alpha\beta}e^{-L_a \beta}}+\frac{e^{(i\alpha+L_a
)(t-\beta)}}{1-e^{-i\alpha\beta}e^{-L_a
\beta}})+\Theta(-t)\frac{1}{2}(-\frac{e^{(i\alpha-L_a
)(t+\beta)}}{1-e^{i\alpha\beta}e^{-L_a \beta}}+\frac{e^{(i\alpha+L_a
)t}}{1-e^{-i\alpha\beta}e^{-L_a \beta}})}

\subsec{Two Loops}
In this section, we will compute the coefficients $C_{1,1,-2}$ from the two loop diagrams. The two loop diagrams contributing to the effective potential
are shown in  figure \diagrams. In particular, we will only need to extract the coefficients multiplying $trUtrUtrU^{\dagger2}+trU^{\dagger}trU^{\dagger}trU^{2}$ in each diagram. Fortunately, these coefficients turn out to be non-divergent. No regularizations are required for them.
The two loop diagrams can be computed to give\

\noindent 2a:
\eqn\twoa{\eqalign{F_{2a}=\sum_{m's,j_{\gamma}}-\frac{\beta g_{YM}^{2}}{2}(D^{\alpha\beta\gamma}D^{\bar{\alpha}\bar{\beta}\bar{\gamma}}-D^{\alpha\bar{\alpha}\gamma}D^{\beta\bar{\beta}\bar{\gamma}})\triangle_{j_{\alpha}}(0,\alpha _{ab})\triangle_{j_{\beta}}(0,\alpha _{bc})
}}
2b:
\eqn\twob{\eqalign{F_{2b}=\sum_{m's}\frac{\beta g_{YM}^{2}}{2}\int dt
\triangle_{j_{\alpha}}^{ab,bc}(t)\triangle_{j_{\beta}}^{cb,de}(t)\triangle_{j_{\gamma}}^{ed,ba}(t)(-E^{\alpha\beta\gamma}E^{\bar{\alpha}\bar{\beta}\bar{\gamma}}+2E^{\alpha\beta\gamma}E^{\bar{\beta}\bar{\alpha}\bar{\gamma}})
}}
2c:
\eqn\twoc{\eqalign{F_{2c}=\sum_{m's} \beta g_{YM}^{2}\frac{D^{\alpha\beta\gamma}D^{\bar{\alpha}\bar{\beta}\bar{\gamma}}}{j_{\gamma}(j_{\gamma}+1)}(D_{t}\triangle_{j_{\alpha}}(0,\alpha _{ab})D_{t}\triangle_{j_{\beta}}(0,\alpha _{ac})+j_{\beta}(j_{\beta}+1)\triangle_{j_{\alpha}}(0,\alpha _{ab})\triangle_{j_{\beta}}(0,\alpha _{ac}))
}}
%In 2c, I am not able to perform the sum over $j_{\gamma}$ explicitly.\%
We have used the notation $\triangle_{j_{\alpha}}(0,\alpha _{ab})$
to signal that the propagator participates in both the a and b index
loop in the sense $\alpha^{a } \otimes 1^{ b } - 1^{a } \otimes
\alpha^{b }$. We isolate the coefficient of $trUtrUtrU^{\dagger2}$ in
each of the above and get:\

\noindent 2a:
\eqn\twoaa{\beta g_{{YM}^{2}}\sum_{j_{\alpha},j_{\beta}}\frac{(2j_{\alpha}+1)(2j_{\beta}+1)}{16\pi^{2}}\frac{e^{-\beta(L_{\alpha}+L_{\beta})}+e^{-\beta(L_{\alpha}+2L_{\beta})}+e^{-\beta(2L_{\alpha}+L_{\beta})}}{4(j_{\alpha}(j_{\alpha}+1)(j_{\beta}(j_{\beta}+1))^{1/2}}}
\noindent 2b:
\eqn\twoba{\eqalign{&\beta\frac{g_{YM}^{2}}{2}\sum_{j_{\alpha},j_{\beta},j_{\gamma}}(-\tilde{A}^{2}(j_{\beta},j_{\gamma},j_{\alpha})-2\tilde{A}(j_{\beta},j_{\gamma},j_{\alpha})\tilde{A}(j_{\alpha},j_{\gamma},j_{\beta}))
\frac{1}{8L_{\alpha}L_{\beta}L_{\gamma}}[\frac{1}{L_{\alpha}+L_{\beta}+L_{\gamma}}\cr
(&e^{-(L_{\beta}+2L_{\gamma})\beta}+e^{-(L_{\alpha}+2L_{\beta})\beta}+e^{-(L_{\gamma}+2L_{\alpha})\beta}+e^{-(2L_{\alpha}+L_{\beta})\beta}+e^{-(2L_{\beta}+L_{\gamma})\beta}+e^{-(2L_{\gamma}+L_{\alpha})\beta})\cr
+&\frac{(e^{-(L_{\beta}+2L_{\gamma})\beta}-e^{-(L_{\alpha}+L_{\gamma})\beta})}{L_{\alpha}-L_{\beta}-L_{\gamma}}+\frac{(e^{-(2L_{\alpha}+L_{\gamma})\beta}-e^{-(L_{\alpha}+L_{\beta})\beta})}{-L_{\alpha}+L_{\beta}-L_{\gamma}}\cr
+&\frac{(e^{-(L_{\alpha}+2L_{\beta})\beta}-e^{-(L_{\beta}+L_{\gamma})\beta})}{-L_{\alpha}-L_{\beta}+L_{\gamma}}+\frac{(e^{-(L_{\alpha}+L_{\gamma})\beta}-e^{-(2L_{\alpha}+L_{\beta})\beta})}{L_{\alpha}+L_{\beta}-L_{\gamma}}\cr
+&\frac{(e^{-(L_{\beta}+L_{\gamma})\beta}-e^{-(L_{\alpha}+2L_{\gamma})\beta})}{L_{\alpha}-L_{\beta}+L_{\gamma}}+\frac{(e^{-(L_{\alpha}+L_{\beta})\beta}-e^{-(2L_{\beta}+L_{\gamma})\beta})}{-L_{\alpha}+L_{\beta}+L_{\gamma}}]}}
\noindent 2c:
\eqn\twoca{\eqalign{\beta g_{YM}^{2}\sum_{j_{\alpha},j_{\beta},j_{\gamma}}&\frac{R_{1}(j_{\alpha},j_{\beta},j_{\gamma})}{j_{\gamma}(j_{\gamma}+1)}[\frac{1}{4}[(e^{-\beta(L_{\alpha}+L_{\beta})}-e^{-\beta(L_{\alpha}+2L_{\beta})}-e^{-\beta(2L_{\alpha}+L_{\beta})})\cr+&(\frac{j_{\beta}(j_{\beta}+1)}{j_{\alpha}(j_{\alpha}+1)})^{1/2}(e^{-\beta(L_{\alpha}+L_{\beta})}+e^{-\beta(L_{\alpha}+2L_{\beta})}+e^{-\beta(2L_{\alpha}+L_{\beta})})]]}}
To obtain the above expressions, we have used several properties of the scalar and vector spherical harmonics. These properties as well as the definitions of $\tilde{A}(j_{\beta},j_{\gamma},j_{\alpha})$, $R_1 (j_{\beta},j_{\gamma},j_{\alpha})$ are listed in the appendix.
We will numerically evaluate the momentum sums. The results are:

\medskip

{
\offinterlineskip
\tabskip=0pt
\halign{
\vrule height2.75ex depth1.25ex width 0.6pt #\tabskip=2em &
 \hfil $#$ \hfil &\vrule  # &

\hfil $#$ \hfil & #\vrule \tabskip=0pt\cr
\noalign{\hrule height 0.6pt}
& Diagram && Value & \cr
\noalign{\hrule} & 2a &&   0.000609 &\cr & 2b &&  -0.013068  &\cr & 2c &&
0.00346 &\cr
\noalign{\hrule height 0.6pt}}}
\medskip
With these values we obtain:
\eqn\ctwot{
\frac{- \beta_c C_{1,1,-2}(\beta_c )^2}{ ({1 \over 2} -C_{2,-2}(\beta_c )) }= -0.000816.
}

\subsec{Three Loops}
To determine the order of the phase transition we will also need the coefficient $C_{1,1,-1,-1}$ from the three loop diagrams. The three loop diagrams contributing to the effective potential are the
 same as those in 3+1 dimension.
We will now list the expressions for the three loop diagrams. We will use the notation
\eqn\ehat{
\hat{E}^{a b c} \equiv E^{a b c} +E^{ b c a }+E^{c a b} ,
}
which is totally antisymmetric in its indicies.

\noindent 
We find the following expressions for the diagrams
3a:
\eqn\Fthreea{\eqalign{
F_{3a} &= -{\beta g_{YM}^4 \over 2} (D^{\a \g \l} D^{\bal \delta \bl} D^{\b \bd \t}
D^{\bb \bg \bt} - 2 D^{\a \bal \l} D^{\g \delta \bl} D^{\b \bg \t} D^{\bb
\bd \bt} + D^{\a \bal \l} D^{\g \delta \bl} D^{\b \bb \t} D^{\bg \bd \bt})
\cr
& \qquad \int dt \; \Delta_{j_\a} (0, \alpha_{ab})  \Delta_{j_\g} (t,
\alpha_{ca})  \Delta_{j_\d} (t, \alpha_{ac}) ( \Delta_{j_\b} (0,
\alpha_{ad})+ \Delta_{j_\b} (0, \alpha_{dc}) ),
}}
3b:
\eqn\Fthreeb{\eqalign{
F_{3b} =& \beta g_{YM}^4 (D^{\a \g \l} D^{\bal \delta \bl} -
D^{\a \bal \l} D^{\g \delta \bl}) {D^{\bg \b \r} D^{\bd \bb \br}
\over j_\r (j_\r + 1)}  \cr
& \qquad \left\{ \Delta_{j_\a}(0, \alpha_{ab})(D_\tau \Delta_{j_\b}(0,
\alpha_{ad}) + D_\tau \Delta_{j_\b}(0, \alpha_{dc})) \right. \cr
& \qquad \qquad \qquad \int dt \; (D_\tau \Delta_{j_\g}(t, \alpha_{ac})
\Delta_{j_\delta}(t, \alpha_{ca}) - \Delta_{j_\g}(t, \alpha_{ac}) D_\tau
\Delta_{j_\delta}(t, \alpha_{ca})) \cr
& \qquad + \Delta_{j_\a}(0, \alpha_{ab})(\Delta_{j_\b}(0, \alpha_{ad}) +
\Delta_{j_\b}(0, \alpha_{dc})) \cr
& \left. \qquad \qquad \qquad \int dt \; (j_\b (j_\b +1) \Delta_{j_\g}(t,
\alpha_{ac}) \Delta_{j_\delta}(t, \alpha_{ca}) - D_\tau \Delta_{j_\g}(t,
\alpha_{ac}) D_\tau \Delta_{j_\delta}(t, \alpha_{ca})) \right\},
}}
3c:
\eqn\Fthreec{\eqalign{
F_{3c} =& -{\beta g_{YM}^4 \over 2} {D^{\a \g \l} D^{\bal \delta \bl} D^{\bg \b \r}
D^{\bd \bb \br} \over j_\lambda (j_\lambda+1) j_\r (j_\r + 1)} \cr
& \int dt (\Delta_{j_\b}(0, \alpha_{ad})+ \Delta_{j_\b}(0, \alpha_{dc})) \cr
& \qquad \left\{ (j_\a (j_\a+1) j_\b (j_\b+1)+j_\g (j_\g+1)j_\delta (j_\delta+1))
\Delta_{j_\a}(0, \alpha_{ab}) \Delta_{j_\delta}(t, \alpha_{ac})
\Delta_{j_\g}(t, \alpha_{ca})  \right. \cr
& \qquad - j_\beta (j_\beta + 1) D_\tau \Delta_{j_\g}(t, \alpha_{ca})(4 D_\t
\Delta_{j_\a}(0, \alpha_{ab}) \Delta_{j_\delta}(t, \alpha_{ac}) +
2 \Delta_{j_\a}(0, \alpha_{ab}) D_\tau \Delta_{j_\delta}(t, \alpha_{ac})) \cr
& \qquad \left. - 2 j_\delta (j_\delta + 1) \Delta_{j_\a}(0, \alpha_{ab})
\Delta_{j_\delta}(0, \alpha_{ac}) \right\} \cr
& + \int dt (D_\tau \Delta_{j_\b}(0, \alpha_{ad}) + D_\tau
\Delta_{j_\b}(0, \alpha_{dc})) \cr
& \qquad \left\{j_\g (j_\g+1) \Delta_{j_\g}(t, \alpha_{ca})(
4 D_\tau \Delta_{j_\delta}(t, \alpha_{ac}) \Delta_{j_\a}(0, \alpha_{ab}) +
2  \Delta_{j_\delta}(t, \alpha_{ac}) D_\tau \Delta_{j_\a}(0,
\alpha_{ab})) \right. \cr
& \qquad - 2 D_\tau \Delta_{j_\a}(0, \alpha_{ab}) D_\t \Delta_{j_\delta}(t,
\alpha_{ac}) D_\tau \Delta_{j_\g}(t, \alpha_{ca}) \cr
& \qquad \left. - 2 (D_\tau \Delta_{j_\a}(0, \alpha_{ab}) \Delta_{j_\delta}(0,
\alpha_{ac}) + 2 \Delta_{j_\a}(0, \alpha_{ab}) D_\tau \Delta_{j_\delta}(0,
\alpha_{ac})) \right\},
}}
3d:
\eqn\Fthreed{\eqalign{
F_{3d} =& -{\beta g_{YM}^4 \over 4} D^{\a \g \l} D^{\b \delta \bl} D^{\bal \bg \r}
D^{\bb \bd \br} \cr
& \qquad  \int dt \Delta_{j_\alpha} (t, \alpha_{ab})
\Delta_{j_\beta}(t, \alpha_{bc}) (2  \Delta_{j_\g}(t, \alpha_{cd})
\Delta_{j_\delta}(t, \alpha_{da}) + \Delta_{j_\delta}(t, \alpha_{cd})
\Delta_{j_\g}(t, \alpha_{da}) ) \cr
&-{\beta g_{YM}^4 \over 4} D^{\a \b \l} D^{\g \delta \bl} D^{\bg \bb \r}
D^{\bal \bd \br} \cr
& \qquad \int dt \Delta_{j_\alpha} (t, \alpha_{ab})
\Delta_{j_\beta}(t, \alpha_{bc}) (\Delta_{j_\g}(t, \alpha_{cd})
\Delta_{j_\delta}(t, \alpha_{da}) - 4 \Delta_{j_\delta}(t, \alpha_{cd})
\Delta_{j_\g}(t, \alpha_{da}) ) ,
}}
3e:
\eqn\Fthreee{\eqalign{
F_{3e} =& \beta g_{YM}^4 (2D^{\a \delta \l} D^{\b \g \bl} -
D^{\a \b \l} D^{\g \delta \bl} - D^{\a \g \l} D^{\b \delta \bl})
{D^{\bg \bd \r} D^{\bal \bb \br} \over j_\r (j_\r + 1)}  \cr
& \qquad \int dt \left\{ D_\tau \Delta_{j_\a}(t, \alpha_{ba})
\Delta_{j_\b}(t, \alpha_{ad}) \Delta_{j_\gamma}(t, \alpha_{cb})
D_\tau \Delta_{j_\delta}(t, \alpha_{dc})) \right. \cr
& \qquad \qquad \left. - \Delta_{j_\a}(t, \alpha_{ba}) D_\tau
\Delta_{j_\b}(t, \alpha_{ad}) \Delta_{j_\gamma}(t, \alpha_{cb})
D_\tau \Delta_{j_\delta}(t, \alpha_{dc})) \right\},
}}
3f:
\eqn\Fthr{\eqalign{
F_{3f} &=- {\beta g_{YM}^4 \over 2} {D^{\a \g \l} D^{\b \delta \bl} D^{\bal \bg \r}
D^{\bb \bd \br} \over j_\l (j_\l +1) j_\r (j_\r +1)} \cr
&\qquad \left[4 \Delta_{j_\a} (0, \alpha_{ab}) D_\tau \Delta_{j_\delta} (0,
\alpha_{cd}) D_\tau \Delta_{j_\b} (0, \alpha_{da}) \right. \cr
&\qquad \qquad -2 (j_\b (j_\b +1) + j_\delta (j_\delta + 1)) \Delta_{j_\a} (0, \alpha_{ab})
\Delta_{j_\delta} (0, \alpha_{cd}) \Delta_{j_\b} (0, \alpha_{da}) \cr
&\qquad +\int dt \left\{ \Delta_{j_\a} (t, \alpha_{ab}) \Delta_{j_\g} (t,
\alpha_{bc}) \Delta_{j_\delta} (t, \alpha_{cd}) \Delta_{j_\b} (t, \alpha_{da})
\right. \cr
&\qquad \qquad \qquad \qquad j_\gamma (j_\gamma +1)(j_\delta (j_\delta + 1) + j_\b (j_\b +1)) \cr
& \qquad \qquad +2 D_\tau \Delta_{j_\a} (t, \alpha_{ab}) D_\tau
\Delta_{j_\g} (t, \alpha_{bc})D_\tau \Delta_{j_\delta} (t, \alpha_{cd}) D_\tau
\Delta_{j_\b} (t, \alpha_{da}) \cr
& \left. \left. \qquad \qquad -4 j_\gamma (j_\gamma+1)
\Delta_{j_\a} (t, \alpha_{ab}) \Delta_{j_\g} (t, \alpha_{bc}) D_\tau
\Delta_{j_\delta} (t, \alpha_{cd}) D_\tau \Delta_{j_\b} (t, \alpha_{da})
\right\} \right]\cr
&- {\beta g_{YM}^4 \over 2} {D^{\a \b \l} D^{\g \delta \bl} D^{\bal \bg \r}
D^{\bb \bd \br} \over j_\l (j_\l +1) j_\r (j_\r +1)} \cr
&\qquad \left[4 D_\tau \Delta_{j_\a} (0, \alpha_{ab}) D_\tau
\Delta_{j_\g} (0, \alpha_{bc}) \Delta_{j_\b} (0, \alpha_{da}) \right. \cr
&\qquad \qquad -2 j_\a (j_\a +1) \Delta_{j_\a} (0, \alpha_{ab})
\Delta_{j_\g} (0, \alpha_{bc}) \Delta_{j_\b} (0, \alpha_{da}) \cr
&\qquad \qquad -2 \Delta_{j_\a} (0, \alpha_{ab}) D_\tau
\Delta_{j_\g} (0, \alpha_{bc}) D_\tau \Delta_{j_\b} (0, \alpha_{da}) \cr
&\qquad +\int dt \left\{ D_\tau \Delta_{j_\a} (t, \alpha_{ab}) D_\tau
\Delta_{j_\g} (t, \alpha_{bc}) D_\tau  \Delta_{j_\delta} (t, \alpha_{cd})
D_\tau \Delta_{j_\b} (t, \alpha_{da}) \right. \cr
& \qquad \qquad +j_\a (j_\a +1) j_\delta (j_\delta +1) ( \Delta_{j_\a} (t, \alpha_{ab})
\Delta_{j_\g} (t, \alpha_{bc}) \Delta_{j_\delta} (t, \alpha_{cd})
\Delta_{j_\b} (t, \alpha_{da}) }}
$$ \qquad \qquad +2 j_\delta (j_\delta +1) ( \Delta_{j_\a} (t, \alpha_{ab})D_\tau
\Delta_{j_\g} (t, \alpha_{bc}) \Delta_{j_\delta} (t, \alpha_{cd}) D_\tau
\Delta_{j_\b} (t, \alpha_{da}) $$
$$ \left. \left. \qquad \qquad -4 (j_\delta+1)^2 D_\tau
\Delta_{j_\a} (t, \alpha_{ab}) D_\tau \Delta_{j_\g} (t, \alpha_{bc})
\Delta_{j_\delta} (t, \alpha_{cd}) \Delta_{j_\b} (t, \alpha_{da}) \right\} \right],$$
3g:
\eqn\Fthreeg{\eqalign{
F_{3g} =& \beta g_{YM}^4 \hat{E}^{\a \delta \r} \hat{E}^{\g \b \br}(D^{\bal \bg \l}
D^{\bb \bd \bl} - {1 \over 2} D^{\bal \bb \l}
D^{\bg \bd \bl} - {1 \over 2} D^{\bal \bd \l} D^{\bb \bg \bl}) \cr
& \qquad \int dt dt' \Delta_{j_\b}(t', \alpha_{da}) \Delta_{j_\g}(t',
\alpha_{cd}) \Delta_{j_\delta}(t, \alpha_{bc})
\Delta_{j_\a}(t, \alpha_{ab}) \Delta_{j_\r}(t'-t, \alpha_{ac}),
}}
3h:
\eqn\Fthreeh{\eqalign{
F_{3h} =&\beta g_{YM}^4 {1 \over j_\l (j_\l + 1)} D^{\a \g \l} D^{\b \delta \bl}
\hat{E}^{\bal \bg \r} \hat{E}^{\bd \bb \br} \cr
& \qquad \int dt_1 dt_2 \; D_\tau \Delta_{j_\a} (t_1, \alpha_{ab})
\Delta_{j_\g}(t_1, \alpha_{bc}) \Delta_{j_\r}(t_1-t_2, \alpha_{ca}) \cr
& \qquad \qquad \qquad (\Delta_{j_\delta}(t_2, \alpha_{cd})
D_\tau \Delta_{j_\b}(t_2, \alpha_{da}) - D_\tau \Delta_{j_\delta}(t_2,
\alpha_{cd}) \Delta_{j_\b}(t_2, \alpha_{da}))\cr
& + \beta g_{YM}^4 {1 \over j_\l (j_\l + 1)} D^{\a \b \l} D^{\g \delta \bl}
\hat{E}^{\bal \bg \r} \hat{E}^{\bd \bb \br} \cr
& \qquad \int dt_1 dt_2 \; D_\tau \Delta_{j_\a} (t_1, \alpha_{ab})
\Delta_{j_\b}(t_2, \alpha_{da}) \Delta_{j_\r}(t_1-t_2, \alpha_{ca}) \cr
& \qquad \qquad \qquad (\Delta_{j_\delta}(t_2, \alpha_{cd})
D_\tau \Delta_{j_\g}(t_1, \alpha_{bc}) -
D_\tau \Delta_{j_\delta}(t_2, \alpha_{cd}) \Delta_{j_\g}(t_1, \alpha_{bc})),
}}
3i:
\eqn\Fthreei{\eqalign{
F_{3i} =& -{\beta g_{YM}^4 \over 4} \hat{E}^{\a \b \r}
\hat{E}^{\bal \s \bb} \hat{E}^{\bs \delta \g} \hat{E}^{\bd \br \bg} \cr
& \qquad \int dt_1 dt_2 dt_3 \Delta_{j_\a}(t_1-t_2, \alpha_{ab})
\Delta_{j_\b}(t_1-t_2, \alpha_{bc})\Delta_{j_\g}(t_3, \alpha_{cd}) \cr
& \qquad \qquad \qquad \qquad \Delta_{j_\delta}(t_3, \alpha_{da})
\Delta_{j_\r}(t_1-t_3, \alpha_{ca})\Delta_{j_\s}(t_2, \alpha_{ac}),
}}
3j:
\eqn\Fthreej{\eqalign{
F_{3j} =& \beta g_{YM}^4(D^{\a \r \l} D^{\b \br \bl} - D^{\r \br \l} D^{\a \b \bl})
\hat{E}^{\bal \t \s} \hat{E}^{\bb \bs \bt} \cr
& \qquad \int dt dt' \Delta_{j_\r}(0, \alpha_{ab}) \Delta_{j_\b}(t-t',
\alpha_{ac}) \Delta_{j_\s}(t', \alpha_{ad})
\Delta_{j_\a}(t, \alpha_{ca}) \Delta_{j_\t}(t', \alpha_{dc}),
}}
3k:
\eqn\Fthreek{\eqalign{
F_{3k} =& \beta g_{YM}^4 {1 \over j_\lambda (j_\lambda + 1)}
D^{\alpha \beta \lambda} D^{\bal \g \bl} \hat{E}^{\bb \r \s}
\hat{E}^{\bg \bs \br} \cr
& \qquad \int dt_1 dt_2 \; \Delta_{j_\r}(t_1 - t_2, \alpha_{cd})
\Delta_{j_\s}(t_1 - t_2, \alpha_{da}) \cr
& \qquad \qquad \qquad \left\{ 2 D_\tau \Delta_{j_\alpha}(0, \alpha_{ab})
\Delta_{j_\b}(t_1, \alpha_{ac}) D_\tau \Delta_{j_\gamma}(t_2, \alpha_{ca})
\right. \cr
& \qquad \qquad \qquad + \Delta_{j_\alpha}(0, \alpha_{ab}) D_\tau
\Delta_{j_\b}(t_1, \alpha_{ac}) D_\tau \Delta_{j_\gamma} (t_2, \alpha_{ca}) \cr
& \qquad \qquad \qquad \left. -j_\alpha (j_\alpha +1) \Delta_{j_\alpha}(0,
\alpha_{ab}) \Delta_{j_\b}(t_1, \alpha_{ac}) \Delta_{j_\gamma}(t_2,
\alpha_{ca}) \right\},
}}
3l:
\eqn\Fthreel{\eqalign{
F_{3l} =& -{\beta g_{YM}^4 \over 12} \hat{E}^{\a \b \t} \hat{E}^{\bb \g \r}
\hat{E}^{\bg \bal \s} \hat{E}^{\br \bs \bt} \cr
& \qquad \int dt_1 dt_2 dt_3 \Delta_{j_\a}(t_2-t_3, \alpha_{ab})
\Delta_{j_\b}(t_3-t_1, \alpha_{ac})\Delta_{j_\g}(t_1-t_2, \alpha_{ad}) \cr
& \qquad \qquad \qquad \qquad \Delta_{j_\r}(t_1, \alpha_{dc})
\Delta_{j_\s}(t_2, \alpha_{bd})\Delta_{j_\t}(t_3, \alpha_{cb}),
}}
3m:
\eqn\Fthreem{\eqalign{
F_{3m} &= 4\beta g_{YM}^4 {D^{\a \b \l} C^{\bl \bal \r} D^{\g \delta \br}
\hat{E}^{\bb \bg \bd} \over j_\l (j_\l +1) j_\r (j_\r + 1)} \cr
& \qquad \int dt ( D_\tau \Delta_{j_\a} (0, \alpha_{ab})
\Delta_{j_\b} (t, \alpha_{ca}) - \Delta_{j_\a} (0, \alpha_{ab})
D_\tau \Delta_{j_\b} (t, \alpha_{ca}) ) \cr
&\qquad \qquad \qquad ( \Delta_{j_\g} (t, \alpha_{dc})
D_\tau \Delta_{j_\delta} (t, \alpha_{ad}) - D_\tau
\Delta_{j_\g} (t, \alpha_{dc}) \Delta_{j_\delta} (t, \alpha_{ad})) \cr
&+2 \beta g_{YM}^4 {D^{\a \delta \l} C^{\bl \b \r} D^{\g \bd \br}
\hat{E}^{\bal \bg \bb} \over j_\l (j_\l +1) j_\r (j_\r + 1)} \cr
& \qquad \int dt \left\{ D_\tau \Delta_{j_\a} (t, \alpha_{ab})
\Delta_{j_\b} (t, \alpha_{bd}) D_\tau \Delta_{j_\g} (t, \alpha_{da})
\Delta_{j_\delta} (0, \alpha_{ca}) \right. \cr
&\qquad \qquad + 2 D_\tau \Delta_{j_\a} (t, \alpha_{ab})
\Delta_{j_\b} (t, \alpha_{bd}) \Delta_{j_\g} (t, \alpha_{da})
D_\tau \Delta_{j_\delta} (0, \alpha_{ca}) \cr
&\qquad \qquad \left. -j_\delta (j_\delta + 1) \Delta_{j_\a} (t, \alpha_{ab})
\Delta_{j_\b} (t, \alpha_{bd}) \Delta_{j_\g} (t, \alpha_{da})
\Delta_{j_\d} (0, \alpha_{ca}) \right\},
}}
3n:
\eqn\Fthreen{\eqalign{
F_{3n} &=\beta g_{YM}^4 {D^{\a \g \r} D^{\b \bg \s} \over
j_\r (j_\r +1) j_\s (j_\s + 1)} \left(
3 {C^{\br \bal \l} C^{\bl \bb \bs} \over j_\l (j_\l + 1)} +
3 {C^{\br \bb \l} C^{\bl \bal \bs} \over j_\l (j_\l + 1)} +
D^{\bal \bl \r} D^{\bb \l \bs} + D^{\bal \bl \bs} D^{\bb \l \br} \right) \cr
& \qquad \left\{ D_\tau \Delta_{j_\a} (0, \alpha_{cb})
\Delta_{j_\g} (0, \alpha_{ac}) D_\tau \Delta_{j_\b} (0, \alpha_{ad})
\right. \cr
& \qquad \qquad +2 D_\tau \Delta_{j_\a} (0, \alpha_{cb})
D_\tau \Delta_{j_\g} (0, \alpha_{ac}) \Delta_{j_\b} (0, \alpha_{ad}) \cr
& \left. \qquad \qquad - j_\g (j_\g + 1) \Delta_{j_\a} (0, \alpha_{cb})
\Delta_{j_\g} (0, \alpha_{ac}) \Delta_{j_\b} (0, \alpha_{ad}) \right\} \cr
&- \beta g_{YM}^4 {D^{\a \g \r} D^{\b \bg \s} \over
j_\r (j_\r + 1) j_\s (j_\s + 1)} \left( 3 {C^{\br \bal \l} C^{\bl \bb \bs}
\over j_\l (j_\l + 1)} + D^{\bal \bl \r} D^{\bb \l \bs} \right) \cr
& \qquad \left\{ D_\tau \Delta_{j_\a} (0, \alpha_{ab})
\Delta_{j_\g} (0, \alpha_{ac}) D_\tau \Delta_{j_\b} (0, \alpha_{ad})
\right. \cr
& \qquad \qquad +2 D_\tau \Delta_{j_\a} (0, \alpha_{ab})
D_\tau \Delta_{j_\g} (0, \alpha_{ac}) \Delta_{j_\b} (0, \alpha_{ad}) \cr
& \left. \qquad \qquad + j_\g (j_\g + 1) \Delta_{j_\a} (0, \alpha_{ab})
\Delta_{j_\g} (0, \alpha_{ac}) \Delta_{j_\b} (0, \alpha_{ad}) \right\} \cr
&- \beta g_{YM}^4 {D^{\a \g \r} D^{\bal \bg \s}
\over j_\r (j_\r + 1) j_\s (j_\s + 1)} \left(
3 {C^{\br \b \l} C^{\bl \bb \bs} \over j_\l (j_\l + 1)} +
D^{\l \b \br} D^{\bl \bb \bs} \right) \cr
& \qquad \left\{ 2 D_\tau \Delta_{j_\a} (0, \alpha_{ab})
D_\tau \Delta_{j_\g} (0, \alpha_{ad}) \Delta_{j_\b} (0, \alpha_{bc})
\right. \cr
& \left. \qquad \qquad + (j_\a (j_\a + 1) +j_\g (j_\g + 1))
\Delta_{j_\a} (0, \alpha_{ab}) \Delta_{j_\g} (0, \alpha_{ad})
\Delta_{j_\b} (0, \alpha_{bc}) \right\}. \cr
}}
Again, we will only need to extract the coefficients multiplying $trUtrUtrU^{\dagger}trU^{\dagger}$ in each diagram.
 These coefficients turn out to be non-divergent as well. No regularizations are required for them. At the three loop level, the Feynman diagrams have very complicated expressions. However, they all have the following strucure:
\eqn\gp{\sum_{j's, m's } G_{j's,m's} I_{j's}(trU, trU^{\dagger}) )
.} where $G_{j's, m's}$ are the group theory factors coming from the
vertices in each diagram. $I_{j's}(trU, trU^{\dagger})$ come from
the propagators. We will expand $I_{j's}(trU, trU^{\dagger})$ in
powers of $trU$, $trU^{\dagger}$ to extract the coefficients of
$trUtrUtrU^{\dagger}trU^{\dagger}$. Fortunately, the relevant terms in
$I_{j's}(trU, trU^{\dagger})$ are very similar to those in 3+1
dimensions, which have been computed \integrals. To apply
them, we only need to replace the masses of the propagators
$(j_\alpha+1 )^2_{3+1}\rightarrow (j_\alpha (j_\alpha+1))_{2+1}$.
With these we can evaluate the angular momentum sums numerically.
The results are
\medskip
{
\offinterlineskip
\tabskip=0pt
\halign{
\vrule height2.75ex depth1.25ex width 0.6pt #\tabskip=1em &
 \hfil $#$ \hfil &\vrule  # &
 \hfil $#$ \hfil &\vrule  # &
 \hfil $#$ \hfil &\vrule  # &
% \hfil $#$ \hfil &\vrule  # &
 \hfil $#$ \hfil & #\vrule \tabskip=0pt\cr
\noalign{\hrule height 0.6pt}
& Diagram && Value && Diagram  && Value & \cr
\noalign{\hrule}
&3a && -0.000182 &&3b && 0.000008 & \cr
&3c && 0.000432 && 3d && -0.000187 & \cr
&3e && -0.0001396 && 3f&& 0.0001399 & \cr
&3g && 0.0021095 && 3h && 0.00685 & \cr
&3i && -0.005715 && 3j && 0.00197 & \cr
&3k && 0.000779 && 3l && -0.00195 &\cr
&3m && 0.000752 && 3n && 0.001682 &\cr
\noalign{\hrule height 0.6pt}  }}
\medskip
The total three loop contribution is
\eqn\cthree{
 C_{1,1,-1,-1}=0.001237
}
We thus obtain the coefficient $ b_c = 0.00503 > 0$. Thus, we conclude that the deconfinement transition is second order, followed at a slightly higher temperature by a continuous transition in which the eigenvalue distribution develops a gap.

\newsec{High Temperature Limit}

In this section, we consider the behavior of the theory at general values of the spatial radius in the limit where the temperature is very large (compared with either the inverse spatial radius or the gauge coupling). In this limit, the Kaluza-Klein modes on the thermal circle become very massive, and the theory is well described by an effective two-dimensional theory on $S^2$. This theory contains a two-dimensional gauge field together with an adjoint scalar coming from the zero mode of $A_0$ on the thermal circle. As in the more familiar 3+1 dimensional case, this scalar receives a Debye mass at one loop, corresponding to screening of electric charge. In our case, the mass was calculated in \dhoker\ to be
$$
m^2 \sim \lambda T \ln \left({T \over \lambda} \right),
$$
where the logarithm arises from a resummation of infrared divergent diagrams\foot{Strictly speaking, there are no infrared divergences since we are working on a spatial sphere. Nevertheless, this infinite volume result should be valid as long as the geometrical infrared cutoff scale $1/R$ is smaller than the dynamical infrared cutoff scale $m$. In this case, we are still required to resum a large number of (finite) diagrams to get the leading contribution to the mass. For smaller radii, the geometrical infrared cutoff dominates and the effective scalar mass will be given by the one-loop contribution.}. For $T \gg \lambda$, this mass is much larger than the scale $M \sim \sqrt{\lambda_2} = \sqrt{\lambda T}$ associated with the two-dimensional Yang-Mills theory, so the model should be effectively described by pure two-dimensional Yang-Mills theory on $S^2$ (as long as the sphere radius is larger than $m^{-1}$). This theory has a third-order phase transition at $R \sim \lambda_2^{-{1 \over 2}} \gg m^{-1}$ \dk, so we conclude that our phase diagram has an additional phase transition line coming from large temperature along the curve
$$
TR \sim 1 / (\lambda R)  \; .
$$
An important question is whether this transition remains sharp for large but finite values of the temperature or whether it becomes smoothed out, either for any non-infinite value of the temperature (i.e. any finite $A_0$ mass) or below some particular temperature. For high temperatures, the question should correspond to asking about the fate of the phase transition in the two-dimensional theory when the mass of an adjoint scalar is reduced from infinity. Unfortunately, even this question seems difficult to approach since the theory is no longer solvable with the adjoint scalar.

The persistence of a sharp phase transition would be guaranteed if there were some order parameter associated with the transition. In the pure two-dimensional Yang-Mills theory on $S^2$, there is in some sense an order parameter for the phase transition, namely the eigenvalue distribution for a maximal area Wilson loop which divides the sphere into two equal areas.\foot{The precise shape of the loop is unimportant since the theory is invariant under area-preserving diffeomorphisms.} This eigenvalue distribution is gapped for radii less than the critical radius, but ungapped above it \order. Away from infinite temperature (where the full theory is on $S^2 \times S^1$), it will presumably no longer be true that the expectation value of a spatial Wilson loop will depend only on the enclosed area, so the extension of the order parameter to finite temperatures is ambiguous. We could for example choose to focus on the Wilson loop around the equator of the $S^2$, and it will certainly be true that the full phase diagram will divide into regions for which the eigenvalue distribution for this Wilson loop is gapped or ungapped, but it isn't clear that the boundary of this region should correspond to some non-smooth behavior of the free energy away from the infinite temperature limit.

If the high-temperature phase transition does remain sharp, an intriguing possibility is that it connects on to the third order phase transition line that originates at the critical temperature in the small volume limit (as in figure 6b). Indeed, in the limits where we have analytic control, both of these phase transitions are third-order transitions associated with gapping for the eigenvalue distribution of Wilson lines. To investigate whether the two transitions might be the same, one approach would be to study the behavior of the eigenvalue distribution for the Polyakov loop in the vicinity of the high-temperature transition. This distribution should be close to a delta function in this high-temperature regime, but could still be either gapped or ungapped depending on how the eigenvalue distribution falls off away from the peak. A transition between these two possibilities may show up as a change in behavior for the eigenvalue distribution of the massive adjoint scalar in the effective two-dimensional theory, for example from a strictly localized distribution to one with an exponential tail. Unfortunately, we have not been able to determine whether such a transition occurs, so we leave it as a question for future work to determine whether the two third-order transitions are connected.

\newsec{Possible Phase Diagrams}

We have seen that for small sphere volumes, our gauge theory undergoes a second order deconfinement transition followed by a third order gapping transition as the temperature is increased. We would now like to understand in general the simplest possibilities for what happens to this behavior as the spatial volume is increased.

First, we expect of course that the deconfinement transition extends all the way to large volume, where the transition temperature should be of order $\lambda$. The simplest logical possibility is that the qualitative behavior we found is unchanged as we go to large volume. This would mean a second order deconfinement transition at large volume. By the conjecture of Svetitsky and Yaffe \sy, the critical behavior should then be the same as a two-dimensional spin model invariant under the same global symmetry, in this case $Z_{N \to \infty} \sim U(1)$. This corresponds to an XY model, for which the transition should be of Kosterlitz-Thouless type.

However, if the expected extrapolation to large $N$ of the lattice results mentioned in the introduction holds, the infinite volume transition should be first order for large $N$, so the Svetitsky-Yaffe predictions would not apply. In this case, which we will assume for the remainder of the section, there must be a critical sphere radius at which the deconfinement transition changes from second order to first order.

We will now argue that there are two different types of behavior possible at the critical radius. To understand this, consider the effective potential for the eigenvalue distribution evaluated at the deconfinement transition temperature for various values of the radius. Where the transition is second order, the potential at the transition temperature has a global minimum for the uniform eigenvalue distribution, with a vanishing second derivative along some direction. The effective potential develops a negative second derivative along this direction as we go above the transition temperature, and the global minimum smoothly moves away from the uniform eigenvalue distribution. There are two qualitatively different effects that can give rise to a change to first order behavior:

First, the effective potential evaluated at the transition temperature might develop a second global minimum at a point away from the uniform distribution for some value of the radius. In this case, if we move further along the line where the uniform distribution is marginally stable (dotted line in figure 3), this new minimum will (generically) become the global minimum, so we must have a first order transition to this new minimum occurring at some lower temperature. Thus, below this critical radius, the deconfinement transition is second order and follows the boundary along which a local instability develops around the uniform distribution. Above the critical radius, the transition is first order and follows the boundary along which we have two global minima. Generically this line of two minima will not terminate at the critical radius but will continue to smaller radii above the deconfinement temperature. Even at these smaller radii, it represents a phase transition, since below the line the minimum near the uniform distribution should be the global minimum, while above the line, the other minimum will be the global minimum. Thus, we have a triple point at the critical temperature and critical radius separating three distinct phases. If the second minimum is at the boundary of configuration space, the higher temperature transition will correspond to a gapping transition, but in this case a first order one, since the eigenvalue distribution jumps discontinuously. The phase diagram in the vicinity of the triple point is sketched in figure 3.

\fig{Phase diagram in the vicinity of the deconfinement transition when we have a triple point at the critical radius, with sketches of the effective potential in each region. Deconfinement transition switches from second order (dashed line) to first order (solid line). Dotted line is not a phase transition but represents boundary in deconfined phase of region for which local minimum exists at the origin.}{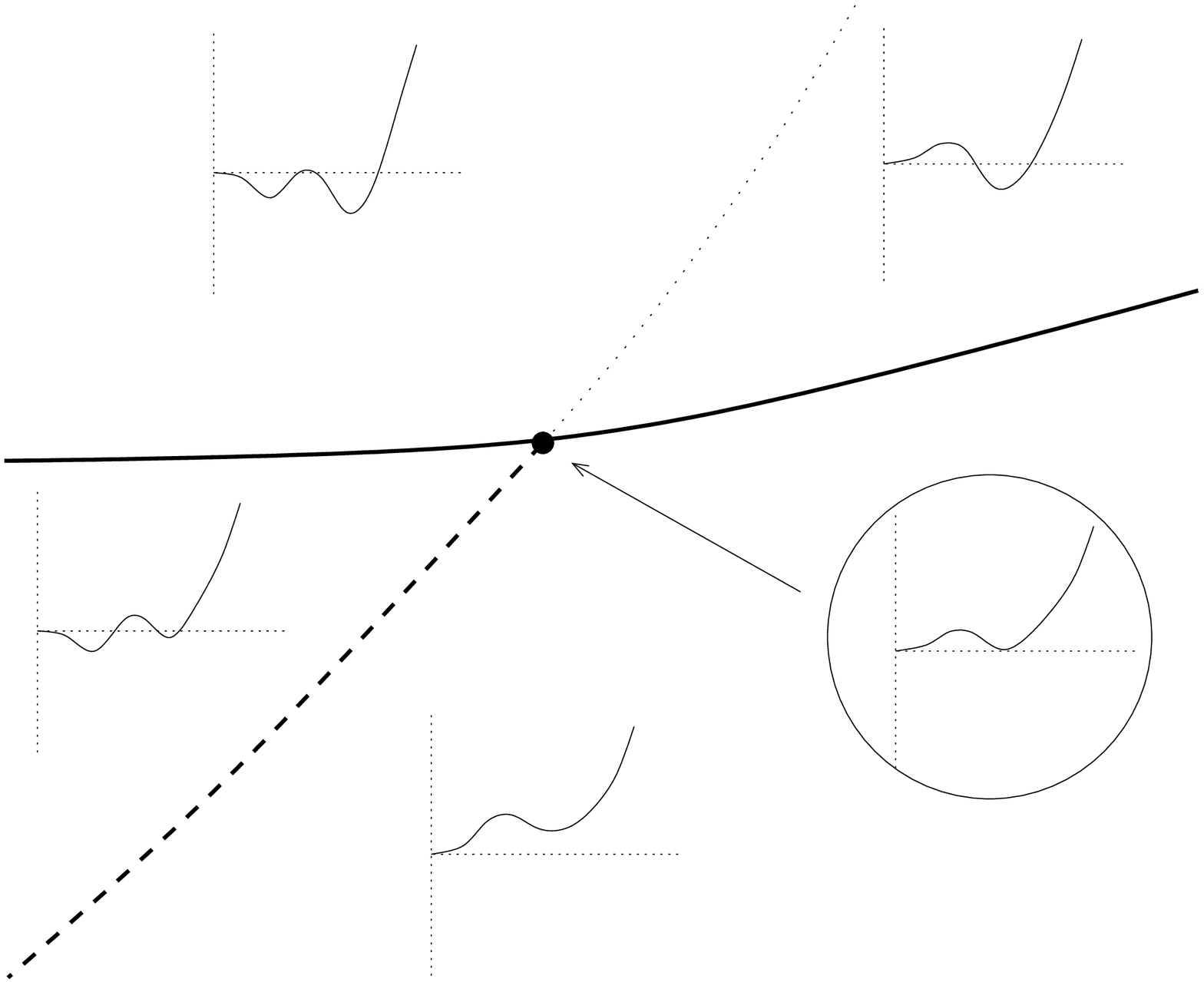}{5.0truein}

We now explain the other possible behavior near the critical radius (depicted in figure 4). If we follow the curve along which a single marginally stable direction exists (with all other directions stable), it may happen that this direction becomes marginally unstable at some point (e.g. if the fourth derivative of the potential in the marginal direction switches from positive to negative). Such a point is known as a tricritical point. In this case, if we continue along the line where we have a single marginal direction, the global minimum of the potential will shift gradually away from the uniform distribution. The deconfinement transition then no longer coincides with the curve along which the uniform distribution becomes locally unstable, but occurs at some lower temperature where the minimum at the origin (which exists everywhere below the marginal stability line) takes the same value as the new nearby minimum. The deconfinement transition corresponds to a jump from the minimum at the origin (i.e. the uniform distribution) to the nearby minimum. It is therefore first order but with a latent heat which vanishes as we approach the tricritical point, where the two minima in the effective potential merge. Unlike the other scenario, there is no other phase boundary emerging from the point at which the deconfinement transition switches behavior, since the line along which we have two equivalent minima simply ends at the tricritical point. In the present case, the first order deconfinement transition either to the left or right of the tricritical point is certainly to an ungapped phase.

\fig{Phase diagram near the critical radius in the case when we have a tricritical point, with sketch of the effective potential in each region. Deconfinement transition switches from second order (dashed line) to first order (solid line). Dotted line is not a phase transition but represents boundary in deconfined phase of region for which local minimum exists at the origin.}{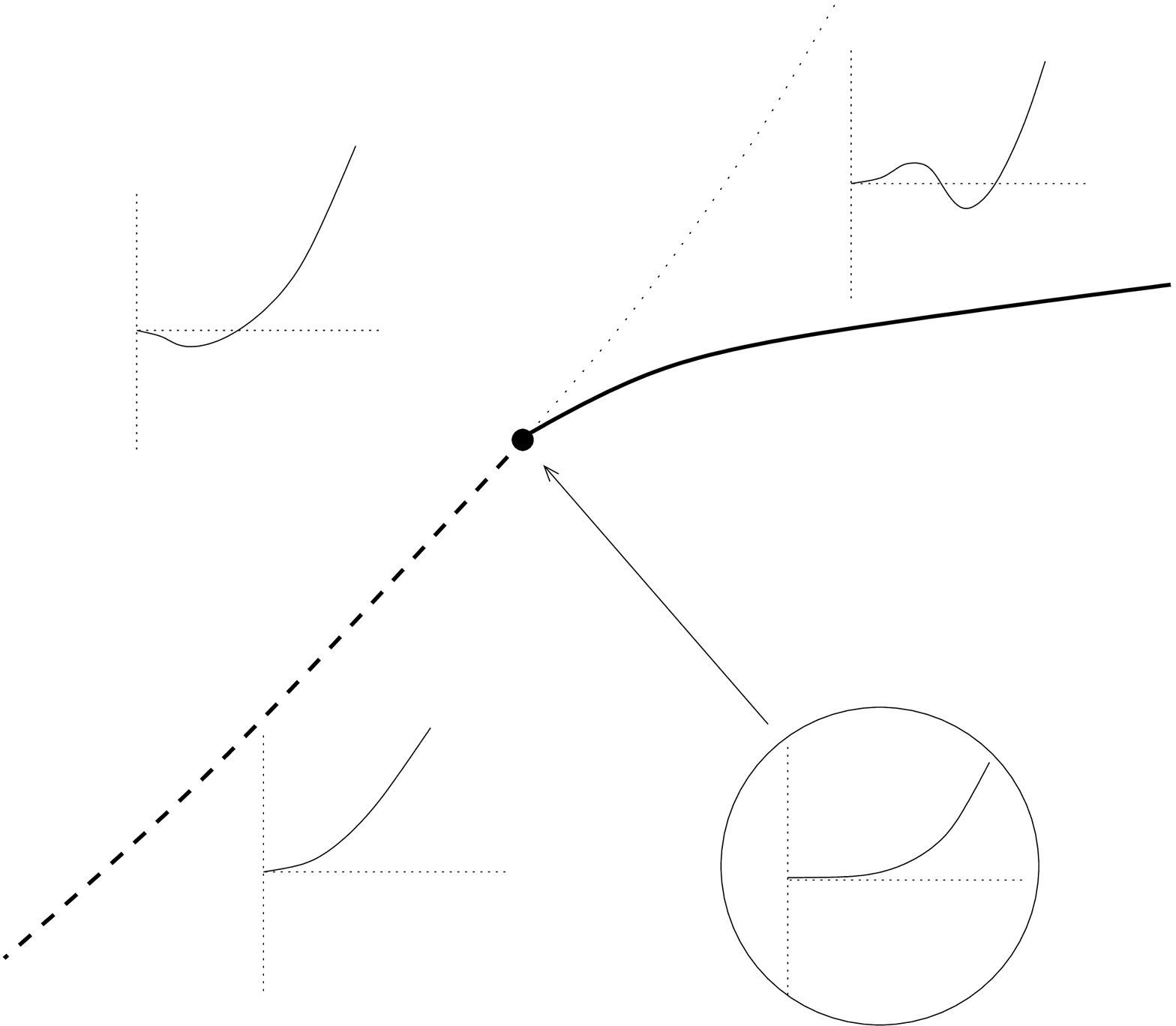}{5.0truein}

Both of these scenarios are realized in a simple toy model for a complex order parameter $w$ with effective potential
$$
S_{eff} = a |w|^2 + b |w|^4 + c |w|^6
$$
and a boundary $|w| = 1$ for the configuration space. The phase diagram for this toy model as a function of the parameters $a$,$b$ is shown in figure 5 for the two cases $c<0$ and $c>0$. In the first case, the switch from second order to first order behavior for the transition corresponds to a triple point, while in the second case, it corresponds to a tricritical point. Note that in the case $c>0$, only the details of the potential near $w=0$ are important, so any higher order terms can be ignored (so long as they do not give rise to a lower minimum). Thus, the behavior of the toy model in the vicinity of the tricritical point should precisely coincide with the behavior of the Yang-Mills theory if we have a tricritical point at the critical radius, since the effective action for $u_1$ will be of this form, with higher order terms that we can ignore.

\fig{Phase diagram for toy model effective potential for $c<0$ and $c>0$ exhibiting the triple point and tricritical point behaviors. Phases I, II, and III correspond to having the global minimum at the origin, in the bulk of the configuration space away from the origin, and at the boundary of configuration space respectively. Solid and dashed lines represent first order and second order phase transitions respectively.}{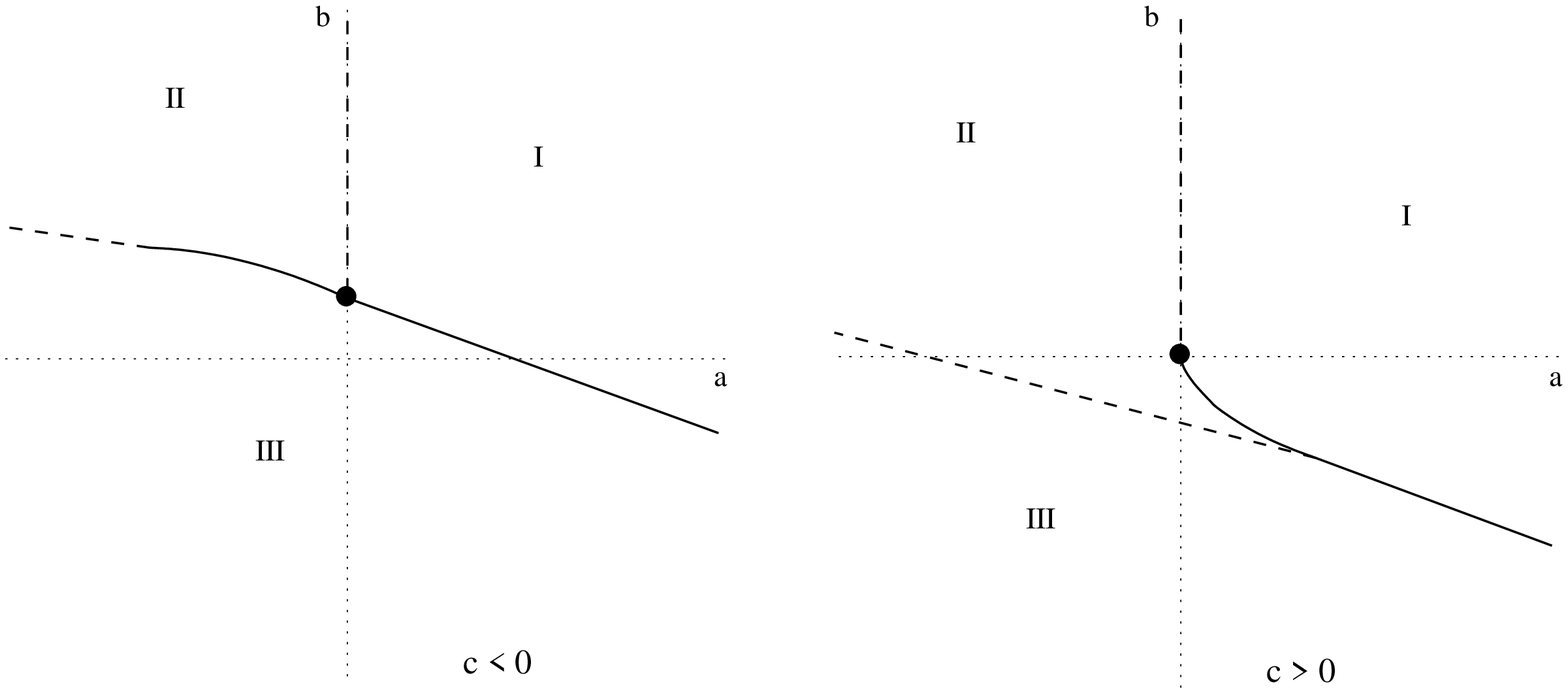}{7truein}

We can now comment on the possible forms for the full phase diagram, assuming that the large-volume transition is first order, so that there exists a change of behavior at some critical radius. If the effective potential is such that we have a triple point, the simplest possibility would be that the additional phase boundary coming from the tricritical point corresponds to a gapping transition, so that this phase boundary would connect with the third order gapping transition emerging from the zero-volume critical temperature. This requires there to be some radius at which the gapping transition switches from third order to first order. In this scenario, the full phase diagram would appear as in figure 6a. The fate of the high-temperature Douglas-Kazakov transition is unclear, though as we have discussed, it is possible that this phase boundary simply ends at some high temperature.

\fig{Simplest possible phase diagrams for large N pure Yang-Mills theory on $S^2$ as a function of sphere radius $R$ and temperature $T$, assuming a first order deconfinement transition at large volume. Solid, dashed, and dotted lines correspond to first, second, and third order transitions respectively.}{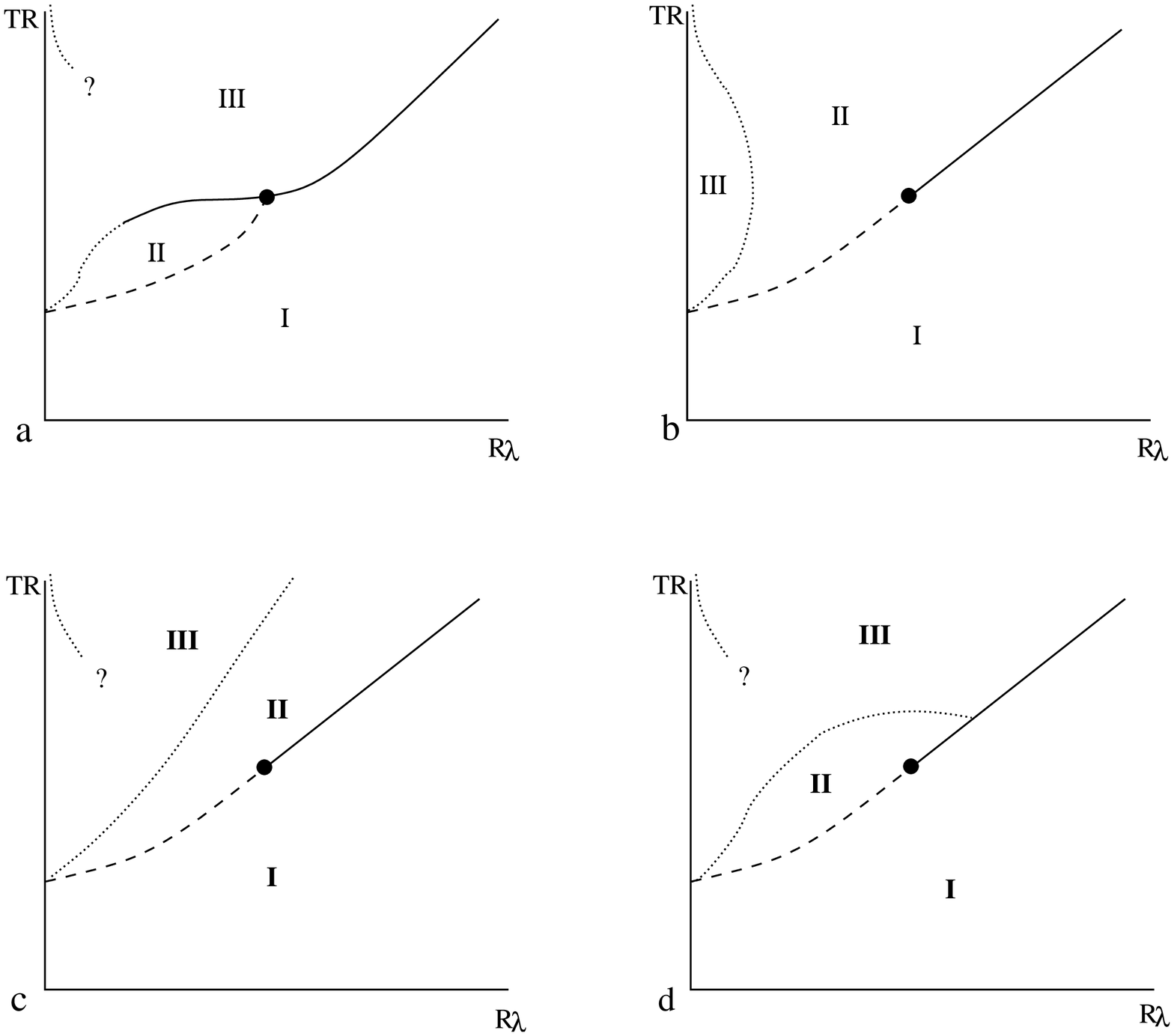}{7truein}

In the case where the effective potential gives rise to a tricritical point, the gapping phase boundary could either extend up to infinite temperature and be absent in the large volume limit (e.g. in the scenario where it connects with the Douglas-Kazakov transition at high temperatures), extend to large volume such that the first order deconfinement transition there would be followed by a gapping transition at some higher temperature, or end somewhere on the deconfinement phase boundary to the right of the tricritical point (as in the toy model for $c>0$). These possibilities are shown in figure 6 b,c, and d respectively.

Further lattice studies should help distinguish between the possibilities in figure 6. In particular, while the distinction between gapped and ungapped eigenvalue distributions strictly exists only in the large N limit, recent studies at relatively large but finite values of $N$ have provided clear suggestions of gapping transitions for eigenvalue distributions of spatial Wilson loops (see, for example \largenlattice). Thus, it should be possible to determine whether the deconfinement transition at large volume is to a gapped or ungapped eigenvalue distribution, and in the latter case, whether there is an additional gapping transition at higher temperature (as in figure 6c). On the other hand, distinguishing between possibilities a) and d) may be difficult, since they differ only at intermediate values of the radius/coupling.

\newsec{Conclusions}

The main result of this paper is that pure large N two-dimensional Yang-Mills theory has a second order deconfinement transition at small spatial volume, with a third-order gapping transition at some higher temperature. This is is a qualitatively different behavior from pure Yang-Mills theory in 3+1 dimensions, and provides the first example of a gauge theory with a single-trace Lagrangian in more than two space-time dimensions for which the deconfinement transition is second order at small volume, and which therefore displays three distinct phases.

If the same behavior exists for some large $N$ theory with a controllable gravity dual, it would be fascinating to understand what the new ungapped phase corresponds to on the gravity side. Generally, the deconfined phase of a large $N$ gauge theory corresponds to a black hole geometry.\foot{The argument \first is that the non-zero expectation value for the Polyakov loop implies that a string worldsheet whose boundary wraps the thermal circle in the associated Euclidean spacetime can have finite area, and therefore the thermal circle must be contractible. In the Lorentzian picture, this is associated with the existence of a horizon.} In ${\cal N} = 4$ SYM theory at strong coupling, the deconfinement transition corresponds on the gravity side to a first-order transition between the original $AdS^5 \times S^5$ spacetime with a thermal gas of supergravity particles to a large black hole spacetime \wittenads. On the other hand, in a theory with an intermediate ungapped phase on the field theory side, there should be a stable intermediate type of black-hole phase on the gravity side smoothly connected to both the no black hole phase and the big black hole phase.

\centerline{\bf Acknowledgements}

We would like to thank Ofer Aharony, Michael Douglas, Igor Klebanov, Joseph Marsano, Shiraz Minwalla, Herbert Neuberger, and Gordon Semenoff for helpful discussions and correspondence. HHS would like to thank Roman Petryk for his help on numerically computing 6-j symbols with large moments. Part of the numerical computations were
done on the Westgrid. The work of MVR was supported in part by the Natural Sciences and Engineering Council of Canada, by the Canada Research Chairs programme, and by the Alfred P. Sloan Foundation. The work of KP was supported by the Foundation of Fundamental Research on Matter (FOM).

%\newsec{Vector Spherical Harmonics}
\appendix{A}{Spherical Harmonics}
In this section we set up our conventions for the vector spherical harmonics $\sva, \svb$.
The definition of a vector spherical harmonics is:
\eqn\vs{
\bar{V}_{JlM}=\sum_{q}V_{JlM}^{q}\hat{e}_{q}=\sum_{m,q} Y_{lm}\hat{e}_{q}(l\,m\,1\,q|\,J\,M)
}
where the $\hat{e}_{q}$ are in the spherical tensor basis:
\eqn\vbasis{\eqalign{
\hat{e}_{+}&=-\frac{\hat{e}_{x}+i\hat{e}_{y}}{2^{1/2}}\cr
\hat{e}_{-}&=\frac{\hat{e}_{x}-i\hat{e}_{y}}{2^{1/2}}\cr
\hat{e}_{0}&=\hat{e}_{z}.
}}
The raising and lowering of the vector index q is given by:
\eqn\rl{
V_{JlM,q}=(-1)^{q}V^{-q}_{JlM}
}
and
\eqn\cc{
\bar{V}_{JlM}^{*}=(-1)^{M+J-l+1}\bar{V}_{Jl-M}.
}
The vector spherical harmonics can be used to expand any well behaved vector fields in $R^3$ and they can be categorized into the following orthonormal basis $\svb$:
\eqn\albasis{\eqalign{
\bar{P}_{JM}&=\frac{1}{(2J+1)^{1/2}}[-(J+1)^{1/2}\bar{V}_{JJ+1M}+J^{1/2}\bar{V}_{JJ-1M}]=\bar{r}Y_{JM}\cr
\bar{B}_{JM}&=\frac{1}{(2J+1)^{1/2}}[J\frac{(J+1)^{1/2}}{r}\bar{V}_{JJ+1M}+(J+1)\frac{J^{1/2}}{r}\bar{V}_{JJ-1M}]=\D Y_{JM}\cr
\bar{C}_{JM}&=-\bar{r}\times \frac{r}{(J(J+1))^{1/2}}\bar{B}_{JM}=-i\bar{V}_{JJM}.
}}
The first of the above does not live in the tangent space of the two sphere, while the second
 do not contribute to the effective action by gauge fixing. We have the following useful expression for $\bar{V}_{JJM}$ following from the last of the above identities:
\eqn\gvs{
V_{JJM,q}=\frac{L_{q}}{(J(J+1))^{1/2}}Y_{JM+q}
}
\eqn\rsops{\eqalign{
L_{+}&=\frac{-1}{2^{1/2}}((J-M)(J+M+1))^{1/2}=-\frac{(-1)^{J-M}}{2}((2J+2)(2J+1)(2J))^{1/2}\left(\matrix{J&J&1\cr M&-M-1&1}\right)\cr
L_{-}&=\frac{1}{2^{1/2}}((J+M)(J-M+1))^{1/2}=\frac{(-1)^{J+M}}{2}((2J+2)(2J+1)(2J))^{1/2}\left(\matrix{J&J&1\cr -M&M-1&1}\right)\cr
L_{0}&=M=\frac{(-1)^{J-M}}{2}((2J+2)(2J+1)(2J))^{1/2}\left(\matrix{J&J&1\cr M&-M&0}\right).
}}
%Also,
%\eqn\r1{
%\D\times \textbf{V}_{JJM}=-i\frac{J}{r}(\frac{J}{2J+1})^{1/2}\textbf{V}_{JJ+1M}+i\frac{J+1}{r}(\frac{J+1}{2J+1})^{1/2}\textbf{V}_{JJ-1M}
%}
%\eqn\r2{
%\frac{r}{(J'(J'+1))^{1/2}}\textbf{P}_{JM}\times \textbf{B}_{J'M'}=-Y_{JM}\textbf{C}_{J'M'},\ \frac{r}{(J(J+1))^{1/2}}\textbf{B}_{JM}\times \textbf{B}_{J'M'}=-\textbf{C}_{JM}\cdot\textbf{B}_{J'M'}\hat{r}
%}
%\eqn\r3{
%\textbf{P}_{JM}\times \textbf{C}_{J'M'}=-\frac{r}{(J'(J'+1))^{1/2}}Y_{JM}\textbf{B}_{J'M'},\ \textbf{C}_{JM}\times \textbf{C}_{J'M'}=\frac{r}{(J(J+1))^{1/2}}\textbf{B}_{JM}\cdot\textbf{C}_{J'M'}\hat{r}
%}

\appendix{B}{Effective Vertices}
As we have seen in section 3, when expended in terms of scalar and vector spherical harmonics, the Lagrangian for pure Yang-Mills theory on $S^2 \times S^1$ contains effective vertices that are integrals of products of spherical harmonics. Here we explicitly compute effective vertices. We will write the results with 3-j symbols:
\eqn\threej{
\left(\matrix{j_{1}&j_{2}&j_{3}\cr m_{1}&m_{2}&m_{3}}\right)=(-1)^{j_{1}-j_{2}-m_{3}}(2j_{3}+1)^{-1/2}(j_{1}\,m_{1}\,j_{2}\,m_{2}|j_{3}\,-m_{3}).
}

A useful formula for sum of products of three 3-j symbols:
\eqn\thrthrj{\eqalign{\sum_{\mu_{1}\mu_{2}\mu_{3}}&(-1)^{l_{1}+l_{2}+l_{3}+\mu_{1}+\mu_{2}+\mu_{3}}\left(\matrix{j_{1}&l_{2}&l_{3}\cr m_{1}&\mu_{2}&-\mu_{3}}\right) \left(\matrix{l_{1}&j_{2}&l_{3}\cr -\mu_{1}&m_{2}&\mu_{3}}\right)\cr
\times &\left(\matrix{l_{1}&l_{2}&j_{3}\cr \mu_{1}&-\mu_{2}&m_{3}}\right)=\left(\matrix{j_{1}&j_{2}&j_{3}\cr m_{1}&m_{2}&m_{3}}\right)\left\{\matrix{j_{1}&j_{2}&j_{3}\cr l_{1}&l_{2}&l_{3}}\right\}.
}}

%\noindent Sum of products of five 3-j symbols:
%\[\sum_{p_{i},q_{i}}=\left(\matrix{J_{2}&j_{2}&\hat{j}_{2}\cr m_{2}&p_{2}&q_{2}}\right)\left(\matrix{J_{3}&j_{3}&\hat{j}_{3}\cr m_{3}&p_{3}&q_{3}}\right)\left(\matrix{J_{1}&j_{1}&\hat{j}_{1}\cr m_{1}&p_{1}&q_{1}}\right)\left(\matrix{j_{1}&j_{2}&j_{3}\cr p_{1}&p_{2}&p_{3}}\right)\left(\matrix{\hat{j}_{1}&\hat{j}_{2}&\hat{j}_{3}\cr q_{1}&q_{2}&q_{3}}\right)\]
%\begin{equation}
%=\left(\matrix{J_{1}&J_{2}&J_{3}\cr m_{1}&m_{2}&m_{3}}\right)\left\{\matrix{J_{1}&J_{2}&J_{3}\cr j_{1}&j_{2}&j_{3}\cr\hat{j}_{1}&\hat{j}_{2}&\hat{j}_{3}}\right\}
%\end{equation}
%\subsection{Integrals}
\noindent $\bullet$ SSS   (given by Gaunt's formula for associated Legendre polynomials) $\svb$:
\eqn\SSS{\eqalign{\int_{S^{2}}Y_{l_{1}m_{1}}Y_{l_{2}m_{2}}Y_{l_{3}m_{3}}&=(\frac{(2l_{1}+1)(2l_{2}+1)(2l_{3}+1)}{4\pi})^{1/2}\left(\matrix{l_{1}&l_{2}&l_{3}\cr0&0&0}\right)\left(\matrix{l_{1}&l_{2}&l_{3}\cr m_{1}&m_{2}&m_{3}}\right)\cr
=&I_{l_{1}l_{2}l_{3}}\left(\matrix{l_{1}&l_{2}&l_{3}\cr m_{1}&m_{2}&m_{3}}\right)
}}
%\subsection{VSV}
\noindent $\bullet$ VSV, The D vertex
\eqn\VSV{\eqalign{
D^{\alpha_{1}\alpha_{2}\alpha_{3}}=\int_{S^{2}}\bar{V}_{l_{1}l_{1}m_{1}}\bar{V}_{l_{2}l_{2}m_{2}}&Y_{l_{3}m_{3}}=\cr
(\frac{(2l_{1}+1)(2l_{2}+1)(2l_{3}+1)}{4\pi})^{1/2}&\frac{l_{3}(l_{3}+1)-l_{1}(l_{1}+1)-l_{2}(l_{2}+1)}{2(l_{1}(l_{1}+1)l_{2}(l_{2}+1))^{1/2}}\cr
\left( \matrix{l_{1}&l_{2}&l_{3}\cr 0&0&0}\right) \left( \matrix{l_{1}&l_{2}&l_{3}\cr m_{1}&m_{2}&m_{3}}\right) &=R^{1/2}_{1}(l_{1},l_{2},l_{3})\left( \matrix{l_{1}&l_{2}&l_{3}\cr m_{1}&m_{2}&m_{3}}\right).
}}
Note $l_{1}+l_{2}+l_{3}$ has to be even for nonzero amplitudes.

\noindent $\bullet$ SVS, The C vertex
\eqn\SVS{\eqalign{C^{\alpha_{3}\alpha_{2}\alpha_{1}}&=\int_{S^{2}}(\D Y_{l_{1}m_{1}})\bar{V}_{l_{2}l_{2}m_{2}}Y_{l_{3}m_{3}} \cr
&=\frac{1}{2r}(\frac{(2l_{1}+1)(2l_{2}+1)(2l_{3}+1)}{4\pi})^{1/2}(\frac{(J+1)(J-2l_{3})(J-2l_{2})(J-2l_{1}+1)}{l_{2}(l_{2}+1)})^{1/2}\cr
&\left(\matrix{l_{1}-1&l_{2}&l_{3}\cr 0&0&0}\right)\left(\matrix{l_{1}&l_{2}&l_{3}\cr m_{1}&m_{2}&m_{3}}\right)=A_{l_{1}l_{2}l_{3}}\left(\matrix{l_{1}&l_{2}&l_{3}\cr m_{1}&m_{2}&m_{3}}\right).
}}
where $J=l_{1}+l_{2}+l_{3}$ and it has to be odd for non-zero amplitudes.\

%\subsection{(curlV$\cdot$r)(V$\times$V$\cdot$r)}
\noindent $\bullet$ (curlV$\cdot$r)(V$\times$V$\cdot$r), The E vertex
\eqn\SVS{\eqalign{E^{\alpha_{3}\alpha_{1}\alpha_{2}}&=\int_{S^{2}}(\D\times V_{J_{3}J_{3}M_{3}}\cdot \hat{r})(V_{J_{1}J_{1}M_{1}}\times V_{J_{2}J_{2}M_{2}}\cdot \hat{r})\cr
&=-(\frac{J_{3}(J_{3}+1)}{J_{1}(J_{1}+1)})^{1/2}A_{J_{1}J_{2}J_{3}}\left(\matrix{J_{1}&J_{2}&J_{3}\cr M_{1}&M_{2}&M_{3}}\right)\cr
&=\tilde{A}_{J_{1}J_{2}J_{3}}\left(\matrix{J_{1}&J_{2}&J_{3}\cr M_{1}&M_{2}&M_{3}}\right).
}}
%This can be seen by first convert $V_{J_{1}J_{1}M_{1}}\times V_{J_{2}J_{2}M_{2}}$ into $\frac{ir}{(J_{1}(J_{1}+1))^{1/2}}\D Y_{J_{1}M_{1}}\cdot V_{J_{2}J_{2}M_{2}}\hat{r}$ using the last identity of (6). Using the completeness of scalar spherical harmonics and (13), we have
%\[(V_{J_{1}J_{1}M_{1}}\times V_{J_{2}J_{2}M_{2}}\cdot \hat{r})=\sum_{J,M} [\frac{ir}{(J_{1}(J_{1}+1))^{1/2}}\int_{S^{2}} \D Y_{J_{1}M_{1}}\cdot V_{J_{2}J_{2}M_{2}}Y_{J-M}(-1)^{M}]Y_{JM}\]
%With the identity (5), wec can again split the rest of the integral into two pieces of the form $V_{010}\cdot V_{J_{3}J_{3}\pm M_{3}}Y_{JM}$ where I have used $\hat{r}=-2(\pi)^{1/2}V_{010}$. Again, using (12), both pieces are proportional to
%\[\left(\matrix{J&0&J_{3}\cr M&0&M_{3}}\right)\]which is nonzero only when $J=J_{3}$, $M=-M_{3}$, and we can explicitly perform the summation over J, M. After evaluating the 6-j symboles and some simplification we obtain (14).
Note since $J=J_{1}+J_{2}+J_{3}$ has to be odd in order for $A_{J_{1}J_{2}J_{3}}$ to be nonzero, $\left(\matrix{J_{1}&J_{2}&J_{3}\cr M_{1}&M_{2}&M_{3}}\right)$ will pick up a negative sign upon interchanging $J_{1}$, $J_{2}$. This suggests $\tilde{A}_{J_{1}J_{2}J_{3}}$ needs to be symmetric in $J_{1}$, $J_{2}$, which can be checked to be true in our expression.

\appendix{C}{Summation Formulas}
%\noindent useful formulas:
The following identities are useful in computing the two loop
diagrams. \eqn\twola{\sum_{m's}
D^{\alpha\beta\gamma}D^{\bar{\alpha}\bar{\beta}\bar{\gamma}}=R_{1}(j_{\alpha},j_{\beta},j_{\gamma})}
\eqn\twolb{\sum_{m's,j_{\gamma}}
D^{\alpha\beta\gamma}D^{\bar{\alpha}\bar{\beta}\bar{\gamma}}=\frac{(2j_{\alpha}+1)(2j_{\beta}+1)}{8\pi}}
\eqn\twolc{\sum_{m's,j_{\gamma}}
D^{\alpha\bar{\alpha}\gamma}D^{\beta\bar{\beta}\bar{\gamma}}=\frac{(2j_{\alpha}+1)(2j_{\beta}+1)}{4\pi}}
where
\eqn\R{\eqalign{R_{1}(l_{1},l_{2},l_{3})=&(\frac{(2l_{1}+1)(2l_{2}+1)(2l_{3}+1)}{4\pi})\frac{(l_{3}(l_{3}+1)-l_{1}(l_{1}+1)-l_{2}(l_{2}+1))^{2}}{4l_{1}(l_{1}+1)l_{2}(l_{2}+1)}\cr
&\left(\matrix{l_{1}&l_{2}&l_{3}\cr0&0&0}\right)^{2}}}
\eqn\twolEE{\sum_{m's}
E^{\alpha\beta\gamma}E^{\bar{\alpha}\bar{\beta}\bar{\gamma}}=+\tilde{A}^{2}(j_{\beta},j_{\gamma},j_{\alpha})
\sigma(J_{\alpha},J_{\beta},J_{\gamma})} \eqn\twolEEa{\sum_{m's}
E^{\alpha\beta\gamma}E^{\bar{\beta}\bar{\alpha}\bar{\gamma}}=(-1)\tilde{A}(j_{\beta},j_{\gamma},j_{\alpha})\tilde{A}(j_{\alpha},j_{\gamma},j_{\beta})\sigma(J_{\alpha},J_{\beta},J_{\gamma})}
where $\tilde{A}(l_{1},l_{2},l_{3})$ is defined as above. and
$\sigma(J_{\alpha},J_{\beta},J_{\gamma})$, is 1 if its arguments
satisfies the triangle inequality, and is zero otherwise. The
presence of $\sigma(J_{\alpha},J_{\beta},J_{\gamma})$ is just a
reminder that we need to impose the triangle inequality on
$J_{\alpha},J_{\beta},J_{\gamma}$ at each vertex which is obvious
from the left hand side but is somewhat obscured by the summation.

The following identities are useful in computing the three loop diagrams. Again, a hat over the indicies means summing over their cyclic permutations.

\eqn\D{\eqalign{\sum_{m}D^{\alpha\overline{\alpha}\lambda}=-(-1)^{J_{\alpha}}R^{1/2}_{1}(J_{\alpha},J_{\alpha},0)(2J_{\alpha}+1)^{1/2}\delta_{J_{\lambda},0}\delta_{m_{\lambda},0}}}

\eqn\DDa{\eqalign{\sum_{m's}D^{\alpha\beta\gamma}D^{\overline{\alpha\beta}\tau}=  (-1)^{J_{\gamma}}\frac{1}{(2J_{\gamma}+1)}  \delta_{J_{\gamma},J_{\tau}}  R_{1}(\alpha,\beta,\gamma )}}

\eqn\DDb{\eqalign{\sum_{m's}D^{\alpha\gamma\tau}D^{\beta\overline{\gamma\tau}}=  (-1)^{J_{\alpha}+1}\frac{1}{(2J_{\alpha}+1)}  \delta_{J_{\alpha},J_{\beta}}  R_{1}(\alpha,\gamma,\tau )}}

\eqn\CC{\eqalign{\sum_{m's}C^{\alpha\epsilon\gamma}C^{\overline{\gamma\epsilon}\beta}=  (-1)^{J_{\alpha}+1}\frac{1}{(2J_{\alpha}+1)}  \delta_{J_{\alpha},J_{\beta}}  A(\gamma\epsilon\alpha) A(\alpha\epsilon\gamma)}}

\eqn\EE{\eqalign{\sum_{m's}E^{\alpha\gamma\epsilon}E^{\overline{\epsilon\gamma}\beta}= (-1)^{J_{\alpha}}\frac{1}{(2J_{\alpha}+1)}  \delta_{J_{\alpha},J_{\beta}}  \tilde{A}(\gamma\epsilon\alpha) \tilde{A}(\gamma\alpha\epsilon)}}

\eqn\DE{\eqalign{\sum_{m's}D^{\gamma\tau\alpha}E^{\overline{\tau\gamma}\beta}= 0}}

\eqn\CD{\eqalign{\sum_{m's}C^{\alpha\gamma\tau}D^{\overline{\gamma}\beta\overline{\tau}}= 0}}

\eqn\quarDa{\eqalign{\sum_{m's,J_{\lambda},J_{\tau}}D^{\alpha\overline{\alpha}\lambda}&D^{\gamma\epsilon\overline{\lambda}}D^{\overline{\gamma\epsilon\tau}}D^{\beta\overline{\beta}\tau}\cr
&=  (-1)^{J_{\alpha}+J_{\beta}}  \delta_{J_{\gamma},J_{\epsilon}}  (2J_{\alpha}+1)^{1/2}(2J_{\beta}+1)^{1/2}  R^{1/2}_{1}(\alpha,\alpha, 0)R^{1/2}_{1}(\beta,\beta, 0)R_{1}(\gamma,\gamma, 0)\cr
&=\delta_{J_{\gamma},J_{\epsilon}}\frac{1}{16\pi^{2}}(2J_{\alpha}+1)(2J_{\gamma}+1)(2J_{\beta}+1)}}

\eqn\quarDb{{\sum_{m's,J_{\lambda}}D^{\alpha\overline{\alpha}\lambda}D^{\gamma\epsilon\overline{\lambda}}D^{\beta\overline{\gamma}\tau}D^{\overline{\beta\epsilon\tau}}=(-1)^{J_{\alpha}+J_{\gamma}} \delta_{J_{\gamma},J_{\epsilon}}  \frac{(2J_{\alpha}+1)^{1/2}}{(2J_{\gamma}+1)^{1/2}} R^{1/2}_{1}(\alpha,\alpha,0)R^{1/2}_{1}(\gamma,\gamma, 0)R_{1}(\beta,\gamma,\tau )}}

\eqn\quarDc{\eqalign{\sum_{m's}D^{\alpha\gamma\lambda}D^{\overline{\alpha}\epsilon\overline{\lambda}}D^{\beta\overline{\epsilon}\tau}D^{\overline{\beta\gamma\tau}}  = \delta_{J_{\gamma},J_{\epsilon}} \frac{1}{(2J_{\gamma}+1)}  R_{1}(\alpha,\gamma,\lambda )R_{1}(\beta,\gamma,\tau )}}

\eqn\quarDd{\eqalign{\sum_{m's}D^{\alpha\gamma\lambda}D^{\overline{\alpha}\overline{\gamma}\tau}D^{\epsilon\beta\overline{\lambda}}D^{\overline{\epsilon\beta\tau}}  = \delta_{J_{\lambda},J_{\tau}} \frac{1}{(2J_{\lambda}+1)}  R_{1}(\alpha,\gamma,\lambda )R_{1}(\epsilon,\beta,\lambda )}}

\eqn\quarDe{\eqalign{\sum_{m's}D^{\alpha\beta\lambda}D^{\gamma\epsilon\overline{\lambda}}&D^{\overline{\beta\epsilon}\tau}D^{\overline{\alpha\gamma\tau}}  = -1^{\sum_{j's}}  R^{1/2}_{1}(\alpha,\beta,\lambda) R^{1/2}_{1}(\gamma,\epsilon,\lambda ) R^{1/2}_{1}(\beta,\epsilon,\tau )R^{1/2}_{1}(\alpha,\gamma,\tau )\cr
           & \left\{\matrix{J_{\gamma}&J_{\alpha}&J_{\tau}\cr J_{\beta}&J_{\epsilon}&J_{\lambda}}\right\}}}

\eqn\quaDquaEa{\eqalign{\sum_{m's,J_{\lambda}}D^{\rho\overline{\rho}\lambda}D^{\alpha\beta\overline{\lambda}}\hat{E}^{\overline{\alpha}\tau\sigma}\hat{E}^{\overline{\beta\sigma\tau}}  &= -(-1)^{J_{\alpha}+J_{\rho}} \delta_{J_{\alpha},J_{\beta}} \frac{(2J_{\rho}+1)^{1/2}}{(2J_{\alpha}+1)^{1/2}}  R^{1/2}_{1}(\alpha,\alpha,0)R^{1/2}_{1}(\rho,\rho,0)\cr
&\tilde{A}_{\hat{\tau\sigma\alpha}}\tilde{A}_{\hat{\sigma\tau\alpha}}}}

\eqn\quaDquaC{\eqalign{\sum_{m's}C^{\sigma\gamma\rho}C^{\overline{\rho\gamma\tau}}D^{\alpha\beta\overline{\sigma}}D^{\overline{\alpha\beta}\tau}  = \delta_{J_{\alpha},J_{\tau}} \frac{-1}{(2J_{\tau}+1)} A_{\rho\gamma\tau}A_{\tau\gamma\rho}R_{1}(\alpha,\beta,\tau )}}

\eqn\quaDquaEb{\eqalign{\sum_{m's}D^{\alpha\beta\lambda}D^{\overline{\alpha}\gamma\overline{\lambda}}\hat{E}^{\overline{\beta}\epsilon\rho}\hat{E}^{\overline{\epsilon\gamma\rho}}  = -\delta_{J_{\gamma},J_{\beta}} \frac{1}{(2J_{\gamma}+1)} \tilde{A}_{\hat{\epsilon\rho\beta}}\tilde{A}_{\hat{\epsilon\gamma\rho}}R_{1}(\alpha,\gamma,\lambda )}}

\eqn\EEEEa{\eqalign{\sum_{m's}\hat{E}^{\gamma\alpha\rho}\hat{E}^{\overline{\rho\alpha\epsilon}}\hat{E}^{\overline{\gamma}\tau\beta}\hat{E}^{\overline{\beta\tau}\epsilon} = \delta_{J_{\gamma},J_{\epsilon}} \frac{1}{(2J_{\gamma}+1)} \tilde{A}_{\hat{\alpha\rho\gamma}}\tilde{A}_{\hat{\alpha\epsilon\rho}}\tilde{A}_{\hat{\tau\beta\gamma}}\tilde{A}_{\hat{\tau\epsilon\beta}}}}

\eqn\quaDquaEc{\eqalign{\sum_{m's}D^{\alpha\gamma\rho}D^{\beta\epsilon\overline{\rho}}\hat{E}^{\overline{\alpha\beta}\tau}\hat{E}^{\overline{\epsilon\gamma\tau}}  =  -(-1)^{\sum_{j's}}  R^{1/2}_{1}(\alpha,\gamma,\rho)R^{1/2}_{1}(\beta,\epsilon,\rho)\tilde{A}_{\hat{\beta\tau\alpha}}\tilde{A}_{\hat{\gamma\tau\epsilon}} \left\{\matrix{J_{\alpha}&J_{\gamma}&J_{\rho}\cr J_{\epsilon}&J_{\beta}&J_{\tau}}\right\}}}

\eqn\EEEEb{\eqalign{\sum_{m's}\hat{E}^{\rho\sigma\tau}\hat{E}^{\overline{\alpha}\beta\overline{\tau}}\hat{E}^{\overline{\beta}\gamma\overline{\rho}}\hat{E}^{\overline{\gamma}\alpha\overline{\sigma}} =  -1^{\sum_{j's}} \tilde{A}_{\hat{\sigma\tau\rho}}\tilde{A}_{\hat{\beta\tau\alpha}}\tilde{A}_{\hat{\gamma\rho\beta}}\tilde{A}_{\hat{\alpha\sigma\gamma}} \left\{\matrix{J_{\rho}&J_{\sigma}&J_{\tau}\cr J_{\alpha}&J_{\beta}&J_{\gamma}}\right\}}}

\eqn\quaDquaCa{\eqalign{\sum_{m's}D^{\alpha\gamma\rho}D^{\beta\overline{\gamma}\sigma}C^{\overline{\rho\beta}\tau}C^{\overline{\tau\alpha\sigma}} = -1^{\sum_{j's}}  R^{1/2}_{1}(\alpha,\gamma,\rho)R^{1/2}_{1}(\beta,\gamma,\sigma)A_{\tau\beta\rho}A_{\sigma\alpha\tau} \left\{\matrix{J_{\alpha}&J_{\gamma}&J_{\rho}\cr J_{\beta}&J_{\tau}&J_{\sigma}}\right\}}}

\eqn\ECquaD{\eqalign{\sum_{m's}\hat{E}^{\alpha\beta\gamma}C^{\rho\overline{\gamma}\sigma}D^{\overline{\beta}\tau\overline{\sigma}}D^{\overline{\alpha\tau\rho}} = -(-1)^{\sum_{j's}}\tilde{A}_{\hat{\beta\gamma\alpha}}A_{\sigma\gamma\rho} R^{1/2}_{1}(\beta,\tau,\sigma)R^{1/2}_{1}(\alpha,\tau,\rho) \left\{\matrix{J_{\beta}&J_{\gamma}&J_{\alpha}\cr J_{\rho}&J_{\tau}&J_{\sigma}}\right\}}}

\noindent expressions that are zero :

\noindent in 3h:
\eqn\quaDquaEd{\sum_{m's}D^{\alpha\gamma\lambda}D^{\beta\epsilon\overline{\lambda}}\hat{E}^{\overline{\alpha\gamma}\rho}\hat{E}^{\overline{\epsilon}\beta\overline{\rho}}  =  \delta_{J_{\lambda},J_{\rho}}\frac{1}{(2J_{\lambda}+1)}  R^{1/2}_{1}(\alpha,\gamma,\lambda)R^{1/2}_{1}(\beta,\epsilon,\lambda)  \tilde{A}_{\hat{\gamma\lambda\alpha}}\tilde{A}_{\hat{\beta\lambda\epsilon}}=0}

\noindent in 3m, 3g:
\eqn\quaDCE{\sum_{m's}D^{\alpha\beta\lambda}C^{\overline{\lambda\alpha}\rho}D^{\gamma\epsilon\overline{\rho}}\hat{E}^{\overline{\beta\gamma\epsilon}} = \delta_{J_{\beta},J_{\rho}}\frac{1}{(2J_{\rho}+1)}  R^{1/2}_{1}(\alpha,\rho,\lambda )R^{1/2}_{1}(\gamma,\epsilon,\rho )  A_{\rho\alpha\lambda }\tilde{A}_{\hat{\gamma\epsilon\rho}}=0}

We can use the following to simplify the above expressions:
\eqn\spa{\sum_{l_{3}}R_{1}(l_{1},l_{2},l_{3})=\frac{(2l_{1}+1)(2l_{2}+1)}{8\pi}}
\eqn\spb{R^{1/2}_{1}(l_{1},l_{1},0)=(-1)^{l_{1}+1}\frac{(2l_{1}+1)^{1/2}}{2 \pi^{1/2}}}

\listrefs

\end